\documentclass[sigconf]{acmart} 

\usepackage{tabularray}
\usepackage{pdfpages}
\usepackage{enumitem}
\usepackage{soul}
\AtBeginDocument{%
  \providecommand\BibTeX{{%
    \normalfont B\kern-0.5em{\scshape i\kern-0.25em b}\kern-0.8em\TeX}}}

\setcopyright{acmlicensed}
\copyrightyear{2018}
\acmYear{2018}
\acmDOI{XXXXXXX.XXXXXXX}

\acmConference[CHI '25]{\textbf{Under Revise and Resubmit}}{June 03--05,
  2018}{Woodstock, NY}
\acmISBN{978-1-4503-XXXX-X/18/06}

\copyrightyear{2025}
\acmYear{2025}
\setcopyright{cc}
\setcctype{by}
\acmConference[CHI '25]{CHI Conference on Human Factors in Computing Systems}{April 26-May 1, 2025}{Yokohama, Japan}
\acmBooktitle{CHI Conference on Human Factors in Computing Systems (CHI '25), April 26-May 1, 2025, Yokohama, Japan}\acmDOI{10.1145/3706598.3713799}
\acmISBN{979-8-4007-1394-1/25/04}

\begin{document}

\title{Governance of Generative AI in Creative Work: Consent, Credit, Compensation, and Beyond}
%

\author{Lin Kyi}
\affiliation{%
   \institution{Max Planck Institute for Security and Privacy}
  \city{Bochum}
  \country{Germany}
 }

 \author{Amruta Mahuli}
 \affiliation{%
   \institution{Max Planck Institute for Security and Privacy}
   \city{Bochum}
   \country{Germany}
 }

 \author{M. Six Silberman}
 \affiliation{%
   \institution{University of Oxford}
   \city{Oxford}
  \country{UK}
 }

 \author{Reuben Binns}
 \affiliation{%
   \institution{University of Oxford}
   \city{Oxford}
   \country{UK}
 }

 \author{Jun Zhao}
 \affiliation{%
   \institution{University of Oxford}
   \city{Oxford}
   \country{UK}
 }

 \author{Asia J. Biega}
 \affiliation{%
   \institution{Max Planck Institute for Security and Privacy}
   \city{Bochum}
   \country{Germany}
 }

\renewcommand{\shortauthors}{Kyi et al.}

\begin{abstract}
Since the emergence of generative AI, creative workers have spoken up about the career-based harms they have experienced arising from this new technology. A common theme in these accounts of harm is that generative AI models are trained on workers' creative output without their consent and without giving credit or compensation to the original creators.

This paper reports findings from 20 interviews with creative workers in three domains: visual art and design, writing, and programming. We investigate the gaps between current AI governance strategies, what creative workers want out of generative AI governance, and the nuanced role of creative workers' \textit{consent, compensation} and \textit{credit} for training AI models on their work. Finally, we make recommendations for how generative AI can be governed and how operators of generative AI systems might more ethically train models on creative output in the future.
\end{abstract}

\ccsdesc[500]{Human-centered computing~Empirical studies in HCI}
\ccsdesc[500]{Social and professional topics~Computing / technology policy}
\ccsdesc[500]{Computing methodologies~Artificial intelligence}
\ccsdesc[300]{Security and privacy~Privacy protections}

\keywords{3 Cs (Consent, Credit, and Compensation), AI governance, AI regulation, Generative AI, Creative work, Knowledge work}



\maketitle

\section{Introduction}
The emergence of generative AI has posed a threat to creative professionals' careers, a particular harm being that it has been known to train on previously-made works~\cite{jiang2023ai, inie2023designing, heikkila2022artist}. Creative work usually has copyright protections~\cite{lee2023talkin}, but generative AI systems have been known to sometimes train their models on copyrighted content. It has been alleged, for example, that GitHub Copilot was trained on copyrighted code~\cite{githubcopilot} and that ChatGPT was trained on copyrighted articles from the \textit{New York Times}~\cite{NYT_openai}. Although AI companies such as Open AI insist that the data they use to train their models are publicly available~\cite{public_data_openAI}, the public availability of the data does not necessarily make it legal to use as training input. Additionally, creative workers have expressed that this data was collected and used without the consent of the original creators, without crediting them, and without compensating them. This has led to several recent lawsuits against generative AI companies~\cite{NYT_openai, nvdia_sued}, identification of potential violations against terms of service~\cite{youtube_openAI}, and protests and strikes from creative workers~\cite{hollywood_actors, hollywood_writers}\footnote{We refer to our participants as ``creative professionals'' and ``creative workers'' rather than more specific terms such as ``artist'' because the former terms are more inclusive, capturing those who work professionally in fields not always considered `artistic' such as programming. We define `creative professionals' as those who are paid to produce creative output for their job, either in a company or as freelancers. We also refer to the work created by creative professionals as ``created output'' or ``creative work'' rather than ``data'' or ``intellectual property'' because we acknowledge that created outputs are human-generated creations that require thought and expertise, and that additional effort is needed to transform these outputs into data or intellectual property~\cite{lee2023talkin}.}.

Copyright infringement is used as a legal reason (or legal ``ground'') by creative workers in holding AI companies accountable, but the core issue is that creative workers want to \textit{protect themselves} from the potential harms that generative AI may present to their careers. Given these harms, the ``3 Cs'' of \textit{consent}, \textit{credit} and \textit{compensation} have been recommended as essential to fair treatment of creative works and workers, both in training generative AI systems and in general~\cite{ccc, 3Cs}.
%
%
In this framework, \textit{consent} refers to collecting freely given and informed consent from the creative worker whose work is used to train AI models. \textit{Credit} explicitly acknowledges the creative worker's contribution to the AI model. Lastly, \textit{compensation} means providing monetary or non-monetary reimbursement to the creative worker for using their work to train an AI model~\cite{ccc, 3Cs}.

In this work, we present the perspectives of creative workers regarding generative AI in their work; we investigate i) whether and how the ``3 Cs'' can serve as a useful orienting framework for generative AI governance in the creative work context; ii) we document what creative workers say they need and want from AI governance to better address the potential harms of generative AI, and iii) we address disruptions to creative work that cannot necessarily be addressed through generative AI governance. AI governance refers to ``the mechanisms and processes that shape and govern AI,'' including norms, ethical frameworks, and regulation efforts designed to address the concerns AI may raise~\cite{butcher2019state}. Effective AI governance is crucial for fostering a safer, more accountable, and more responsible generative AI ecosystem~\cite{cihon2020should}. 

To understand what creative professionals want out of generative AI governance structures and to address potential harms, it is useful (indeed arguably essential) to involve stakeholders, especially those creating the work, in the discussions. To this end, we conducted 20 semi-structured interviews with creative professionals in visual art and design, writing, and programming. All three domains require professionals to create items (i.e., the ``creative works''), which may be tangible or digital and may be novel or generic. Professional creative practice tends to involve using established techniques while allowing flexibility to express or embody a particular style (personal, organizational, or associated with a particular project such as a game or body of work). 



\paragraph{Contributions}
This paper makes three main contributions:

\begin{enumerate}
    \item We found that the role of the ``3 Cs''\textemdash{}consent, credit, and compensation\textemdash{}for protecting creative workers is not as straightforward as the current discourse may suggest. We build on the existing literature by articulating these complexities with insights from our interviewees.
    \item We found a gap between the protections and governance structures that creative workers want around generative AI, and what is currently in place at the governmental level, along with companies, publishers, and platforms that creatives work with. We articulate the ``missing'' protections and structures our interviewees identified.
    \item Based on our findings, we provide recommendations for improving generative AI governance at company, publisher, platform and broader governmental levels to address creative workers' interests, and for \textit{concretely} applying the abstract concepts of the ``3 Cs'' in a manner that responds effectively to workers' substantive concerns.
\end{enumerate}
\section{Background}
Previous works have examined how creative workers are impacted by generative AI~\cite{wu2023not, jiang2023ai} as well as how they use generative AI tools in practice~\cite{inie2023designing, woodruff2024knowledge, 10.1145/3613904.3642133, 10.1145/3613904.3641889}. To protect and promote the welfare of creative workers, the 3 Cs (consent, credit, and compensation) and their underlying principles have previously been recommended, both for generative AI and in general settings~\cite{ccc, 3Cs, im2021yes, EUdataregulations2018, Sprigman, Fisk}. However, there is a lack of empirical research on whether this framework would be helpful to creatives. More investigation is required of workers' perspectives on the governance of generative AI\textemdash{}in both its training and its use\textemdash{}in the context of creative work. 

\subsection{The Current State of AI Governance}
AI governance is relatively unorganized due to the various stakeholders, domains, and competing interests involved~\cite{butcher2019state}. Due to the breadth of potential concerns that AI governance needs to address, it has been argued that AI cannot be extensively regulated because it consists of several technologies, presenting different risks across different domains~\cite{cihon2020should, stone2022artificial}. AI is governed at multiple levels: privately by individual companies through voluntary self-regulatory schemes and standards, publicly by government regulation, and through the third sector, through academic research, pressure from trade unions, or NGO initiatives. 

\paragraph{Private Sector}
At the private level, several tech companies involved in developing, researching, and releasing AI technologies have started initiatives to address concerns that AI presents, such as Google DeepMind's Ethics and Society research unit in 2017~\cite{deepmind}, IBM releasing an open source fairness toolkit to help prevent discrimination and bias in machine learning models~\cite{ibm}, and various companies sharing more information about their AI systems~\cite{butcher2019state}.

\paragraph{Public Sector}
The global public regulation of AI is in its infancy; as of 2023, 31 countries have passed AI legislation, and 13 are currently discussing their AI legislation~\cite{csis}. In 2024, the European Union released their AI Act, which is one of the most comprehensive AI regulations to date, ``ensuring that AI systems respect fundamental rights, safety, and ethical principles and by addressing risks of very powerful and impactful AI models''~\cite{ai_act}. However, many existing legal regimes such as copyright, data protection, labor law, and defamation law, may already be applied to certain AI harms despite not being technology-specific.

Global AI governance is currently fragmented and evolving~\cite{cihon2020should}, and multiple jurisdictions are working to create their own AI governance strategies, requiring them to sometimes cooperate with other jurisdictions for multilateral governance efforts~\cite{iapp}. The AI governance efforts of various countries tend to focus on developing comprehensive AI legislation, including specific use cases, and establishing guidelines and standards~\cite{iapp}.

\paragraph{Third Sector}
At the non-profit level, NGOs, trade unions, civil society groups and research organizations are working on AI governance, particularly governance that impacts their specific area of focus~\cite{butcher2019state}. Examples include universities launching research groups dedicated to studying AI ethics and governance~\cite{butcher2019state}, labor unions negotiating rights through strikes ~\cite{hollywood_actors, hollywood_writers}, and organizations such as the Institute of Electrical and Electronics Engineers (IEEE) establishing ``standards, training and education, certification programs, and more, to empower stakeholders designing, developing, and using AIS (autonomous intelligent systems''~\cite{ieee_ais}.

\subsection{Law and the 3 Cs}
The 3 Cs--\textit{consent, credit,} and \textit{compensation}--were first coined by Monica Boța-Moisin in 2017~\cite{ccc}, prior to the widespread availability of generative AI models, in the context of protecting Indigenous People's cultural property. The phrase has more recently emerged as a rallying cry for creative workers in response to generative AI. However, they are not a concrete framework (e.g., in the sense of an objectively implementable technical standard) nor formally recommended by regulation, but the three principles behind it have been used and recommended in other domains previously, and these three principles are individually important in a range of related domains. 
For example, under the European Union's General Data Protection Regulation~\cite{EUdataregulations2018}, asking users for consent one of the six lawful basis for processing personal data and recommended in a range of settings; more generally, consent has been recommended as a best practice for online interactions to prevent harm~\cite{im2021yes}. Giving creative workers credit for their work has been empirically shown to be highly valuable to them and helps their careers~\cite{Sprigman, Fisk}. Additionally, U.S. copyright law usually assumes that creative workers are compensated, either monetarily or otherwise, when their works are used.\footnote{Legal scholarship has however noted that fans creating derivative work (e.g., ``fan art'') based on copyrighted material may use disclaimers as a method of attribution instead of offering compensation~\cite{tushnet2007payment}.}

The introduction of generative AI models like Stability AI's Stable Diffusion and OpenAI's ChatGPT in November 2022 has resulted in a litany of lawsuits filed by creative workers where, at the core, the demands from creative workers remain the same --- consent, compensation, and credit from these generative AI companies for the use of artists' copyrighted expressive works~\cite{Sarah, Getty, Concord, NYT, Silverman, Dudesy, Lehrman}. When distilled, the grievances in most of these lawsuits boil down to: 
\begin{itemize}
    \item \textit{Consent:} Generative AI companies use creative workers' work for training without their authorization or permission, indicating a lack of consent.
    \item \textit{Credit:} Generative AI companies misappropriate creative workers' names, styles, or likenesses for commercial benefit or fail to attribute credit to creative workers.
    \item \textit{Compensation:} Generative AI companies fail to compensate creative workers or adhere to licensing arrangements.
\end{itemize}

Given that the objective of intellectual property law is to incentivize the generation of more creative work~\cite{USConstitution}, one may assume that these attributes of ensuring consent, compensation, and credit are incentivized and protected. However, there is a gap between the characteristics protected by laws and the attributes valued by creative workers~\cite{Fisk}. The natural law justification of creative workers' rights granted under copyright law stems from John Locke's theory (Lockean theory), which entitles every person to earn the fruits of their labor~\cite{Leaffer}. It acknowledges the need to allow oneself to mix `products of the earth' with one's own labor (a `rational, value-creating activity,')~\cite{Locke} to create new things and earn money needed to maintain one's life and happiness. The value of money, for Locke, is derived from money being a medium of consent that allows or disallows people to use their or others' creative labor~\cite{Locke_1}. Therefore, under this theory, copyright law and its exclusive rights enable creative workers to consent and seek compensation for their creative works. 

However, this theory does not consider credit and its valuation for creative workers. Studies indicate that creative workers are willing to sacrifice economic payments to receive attribution for their work~\cite{Sprigman}. In the U.S., although copyright law related to derivative works, misappropriation law, and the right to publicity prevent others from using one's name, work, style, or likeness without accreditation, none of these laws award attribution as an enforceable right despite calls for this right to attribution to be a legally enforceable term in work contracts~\cite{Fisk}. In most other countries in the world, attribution is strongly protected as part of the creator's `moral rights.' This derives from the French legal tradition of \textit{droit d'auteur} (author's right), which stresses the creator's personal connection to the work, and their moral right to be acknowledged as its source\cite{monta1958concept}.


\subsection{Impacts of Generative AI on Creative Workers}

As a result of generative AI relying on large amounts of data to train its models and generate creative content, many copyright issues have emerged, such as those surrounding authorship, fair use, licensing, and liability, among others~\cite{lee2023talkin, legal_genAI}. This is mainly because generative AI can mimic both the \textit{content} and \textit{style} of creative output; \textit{content} referring to ``the objects and concepts in a piece of art'', and \textit{style} referring to how the art is expressed~\cite{wu2023not}.

Creative workers around the world have spoken up about the harms of generative AI on their work, mentioning issues such as damage to their professional reputation, economic losses, plagiarism, copyright issues, and an overall decrease in creative jobs~\cite{jiang2023ai}. The creative community recognizes the potential harms generative AI can have on their work and has started taking action. For example, the SAG-AFTRA and Writers Guild of America went on strike in 2023 to protest against issues relating to intellectual property and usage of AI~\cite{hollywood_actors, hollywood_writers}.

Generative AI can only mimic the technical aspects of producing creative content, but these re-creations are not based on artistic training or traditions, which are developed over many years of practice to create a style unique to each creative professional~\cite{oritz_art}. This mimicry is especially concerning given that when AI-generated works are framed as being anthropomorphic, users perceive the AI as being more ``human'', which changes how users allocate responsibility~\cite{epstein2020gets}. Increasingly, anthropomorphic descriptions are used to describe AI~\cite{baria2021brain, jiang2023ai}.

Currently, there are no formal protections for creative workers, and therefore creative workers must take action by themselves if they want to prevent their work from being potentially trained on. Watermarking, which is where content is marked to convey information, such as its provenance and authenticity~\cite{watermarking}, has been suggested as another way of protecting creative work. However, empirical research has shown that watermarking is not very effective, and researchers suggest new methods be used~\cite{jiang2023evading}. \citet{shan2023glaze} created \textit{Glaze,} a system that, when applied on training data such as artwork shared online, misleads generative AI models and prevents the models from mimicking the artists' style. The researchers found that \textit{Glaze} was effective at preventing style mimicry~\cite{shan2023glaze}. \citet{shan2024nightshade} have followed this up with \textit{Nightshade}, a prompt-specific poisoning attack that allows creative workers to defend themselves against web crawlers.


In a 2024 survey with creative industry executives, it was found that creative businesses have incorporated generative AI at a higher rate compared to other businesses (25\% of creative businesses compared to 3.9\% of businesses across the economy)~\cite{future_unscripted}. The survey also reports that 75\% of participants believe that generative AI tools would eliminate, reduce, or amalgamate different jobs in the creative industry. The report estimates that by 2026, over 203,000 entertainment industry jobs will be impacted or replaced by generative AI in the US~\cite{future_unscripted}. Furthermore, the Game Developer's 2024 Conference report found similar findings: 50\% of surveyed game developers are working at a studio that uses generative AI tools, and 42\% had ethical concerns about the use of generative AI in their industry~\cite{game_conf}. Due to ethical concerns, 57\% of participants felt that unionization amongst game industry workers was important, but only 23\% reported that they were part of a union or that unionization was discussed at work~\cite{game_conf}. \citet{widder2023s} found that software engineers, including those with more power, felt they lacked power over their own work, leading to issues where they could not resolve ethical concerns.


It is not only ``traditional'' forms of creative work that generative AI impacts. Different forms of work have different chances of replacement by AI, with ``automation-prone'' jobs, such as writing, engineering, and programming, being more likely to be replaced~\cite{felten2021occupational}. The release of AI programming tools such as GitHub Copilot has also impacted software development. \textit{Code churn}, which is where code is changed soon after it is written, is expected ``to double in 2024 compared to its 2021, pre-AI baseline'', indicating that generated code is not of high quality~\cite{copilot}. 

However, there are also potential benefits and future opportunities for collaboration between creative workers and generative AI. Inie et al.'s study found that many creative workers are excited about generative AI, and designing more participatory AI can empower creatives to coexist with generative AI~\cite{inie2023designing}. Generative AI was seen as exciting to some creatives because they expect it might increase their productivity, inspire them, and lead to higher quality work~\cite{inie2023designing}. Additionally, creatives were not as concerned about generative AI, maintaining that AI cannot produce high-quality work indistinguishable from human-made work and that creative work is more complex to create than AI is capable of creating~\cite{inie2023designing}. This finding was echoed by Woodruff et al. who found that knowledge workers do not see generative AI as an immediate threat, viewing it more as a tool for making one's work easier, replacing menial tasks, and requiring human supervision~\cite{woodruff2024knowledge}. Han et al. found that generative AI can be used to improve collaboration within teams~\cite{10.1145/3613904.3642133}. Boucher et al. have speculated that generative AI can help empower creative professionals when generative AI is used to honor creative professionals' feedback and perspectives~\cite{10.1145/3613904.3641889}.
\section{Methods}
We interviewed 20 creative professionals to explore how creative workers viewed generative AI and its governance. Interviews were one hour long and took place on Zoom. Our institutional review board approved our study, and participants were compensated \texteuro30 for their participation. 

\paragraph{Selecting Creative Domains}
We interviewed creative workers in visual art and design, writing, and software development. Our creative domains of interest were informed from previous research in psychology that identified six domains of creativity in a \textit{Creative Behavior Inventory}: mathematics and science, music, fine arts, performing arts, literature, and miscellaneous~\cite{hocevar1979development}. Of these six creative domains, we selected and adapted fine arts (visual art and design), mathematics and science (programming), and literature (writing). We selected these three creative domains because they are most relevant to the HCI and broader computer science community, where design, writing, and programming are common elements of the work and research pipeline.  

\subsection{Designing Interviews with Stakeholder Input}
We wanted to represent the perspectives of creative professionals and ensure the interviews included relevant insights for the creative community. To achieve this, we conducted a pilot study. We contacted three creative professionals from our personal networks in the creative domains of interest to discuss our preliminary interview questions, understand what they wish to know about other creative professionals relating to generative AI, and other considerations to take into account during our interviews. Our pilot participants were informed that their insights would be used to form and adjust our interview questions, and not used in reporting results. The three creative professionals were recruited from the authors' own networks, and all make an income from their creative work, either as full-time employees for a company, or part-time through freelancing; we interviewed two designers and one writer for the pilot study. As some of the research team had professional experience as software developers, they provided insights about questions and perspectives programmers may have for the interviews.

Our pilot participants' backgrounds are described below: 
\begin{itemize}
    \item A full-time mid-career video game designer for a small game studio based in the US with experience working in large video game studios (corresponds to P6).
    \item A part-time freelance creative writer with over 5 years of experience publishing with various magazines and journals, based in the UK (P7).
    \item A full-time junior career UX designer for a small start-up based in France (P11). 
\end{itemize}

Based on the pilot interviews, we adjusted our interview questions, taking into account common themes of interest pilot participants mentioned and what they wished to know about other creatives' experiences with generative AI. 

All three pilot participants were invited again to participate in our study, which they all took part in. We recruited them again because our preliminary interview questions significantly changed based on their feedback, ensuring we would hear new insights from pilot participants during the official interviews.

\subsection{Interview Protocol}
The semi-structured interviews ran for about an hour and took place on Zoom. All interviews were conducted in English by the first author. Our interview protocol can be found in Appendix~\ref{sec:appendixa}. 

Participants were first given a briefing on the study and were informed that we were interested in hearing about their experiences and thoughts of generative AI as creative workers. As we had industry participants whose employers were building and/or using generative AI tools, we assured participants that this study is purely for research purposes, and their identities and companies would be anonymous during the data analysis and reporting of results. If participants agreed to continue, they were asked to sign the consent form and fill in the demographics questionnaire. Afterwards, the interview began, and the researcher turned on the audio recording feature on Zoom for data analysis purposes. 

\subsection{Participants}
\subsubsection{Participant Recruitment}
To recruit participants from specific creative roles, we used snowball sampling and recruiting freelancers on Fiverr\footnote{\url{https://www.fiverr.com/}}, a freelancing platform covering a variety of knowledge work. We aimed to recruit creative workers working in different forms of employment, such as within a company or freelancing, to better understand how employment style impacts creative workers potentially affected by generative AI. Participants had to be over 18, speak English, and be professionally working in our three creative domains of interest; experience with generative AI was not necessary for our study. We stopped recruiting at 20 since we reached saturation, generally reached at 9 to 17 interviews~\cite{hennink2022sample}. 

For snowball sampling, the researchers initially reached out to their personal connections and recruited interested participants. At the end of the interviews, the researchers asked participants to share the study information with their connections who might also be interested in participating in the study. 

To recruit on Fiverr, the first author created a profile and created a \textit{Brief}, which is where a client seeking a service posts a gig and gets matched with potential freelancers. Our brief was posted three times, each time to a different target group of freelancers (art/design, programming, and writing), and described that we were seeking creative professionals for an interview study. Fiverr participants were compensated through the Fiverr platform (to comply with Fiverr's Terms of Service) and we covered the service fees.


\subsubsection{Participant Demographics}
Seven participants worked professionally as artists or designers, six as writers, and seven as developers/engineers. We recruited nine men and eleven women, with an average age of 28.63 years old (one participant did not disclose their age), and had an average of 5.23 years of professional experience. Eight participants were working as freelancers, including a few writers who worked with publishers alongside their freelancing, and 12 as employees of a company. Of those employed by a company, most were working in large companies with over 250 employees~\cite{oecd_business}. Most participants were based in the Global North, with some participants based in the Global South. 
Detailed participant demographics can be found in Table~\ref{tab:demographics}.

\begin{table*}[!h]
\centering
\caption{Participant Demographics}
\begin{tabular}{cccccccc} 
\toprule
              & \textbf{Job title}                                                           & \begin{tabular}[c]{@{}c@{}}\textbf{Form(s)}\\\textbf{of employment}\end{tabular}          & \begin{tabular}[c]{@{}c@{}}\textbf{(If working~in~}\\\textbf{a company)}\\\textbf{Number of~}\\\textbf{employees}\end{tabular} & \begin{tabular}[c]{@{}c@{}}\textbf{Years of}\\\textbf{experience}\end{tabular} & \textbf{Location} & \textbf{Age} & \textbf{Gender}  \\ 
\toprule
\textbf{P1}   & UX designer                                                                  & Company                                                                                       & > 250                                                                                                                           & 4                                                                              & Canada            & 28           & Female           \\ 
\midrule
\textbf{P2}   & \begin{tabular}[c]{@{}c@{}}Software development\\engineer\end{tabular}       & Company                                                                                       & > 250                                                                                                                           & 8                                                                              & USA               & 29           & Male             \\ 
\midrule
\textbf{P3}   & Software engineer                                                            & Company                                                                                       & > 250                                                                                                                           & 6                                                                              & USA               & 27           & Male             \\ 
\midrule
\textbf{P4~}  & Trading associate                                                            & Company                                                                                       & Under 10                                                                                                                       & 4                                                                              & Cayman Islands    & 27           & Male             \\ 
\midrule
\textbf{P5~}  & Copywriter                                                                   & Company                                                                                       & > 250                                                                                                                           & 6                                                                              & Germany           & 30           & Male             \\ 
\midrule
\textbf{P6~}  & \begin{tabular}[c]{@{}c@{}}Principal visual\\development artist\end{tabular} & Company                                                                                       & Between 50~to 249                                                                                                              & 17                                                                             & USA               & 39           & Female           \\ 
\midrule
\textbf{P7~}  & Creative writer                                                              & \begin{tabular}[c]{@{}c@{}}Freelance,\\publishers\end{tabular}                                & N/A                                                                                                                            & 8                                                                              & UK                & 25           & Female           \\ 
\midrule
\textbf{P8}   & Software engineer                                                            & Company                                                                                       & Between 10 to 49                                                                                                               & 3                                                                              & Canada            & 26           & Male             \\ 
\midrule
\textbf{P9}   & \begin{tabular}[c]{@{}c@{}}Graphic designer~\\and illustrator\end{tabular}   & Freelance                                                                                     & N/A                                                                                                                            & 10                                                                             & The Netherlands   & 28           & Female           \\ 
\midrule
\textbf{P10}  & Software engineer                                                            & Company                                                                                       &  > 250                                                                                                                           & 4                                                                              & Canada            & 26           & Male             \\ 
\midrule
\textbf{P11}  & Product designer                                                             & Company                                                                                       & Between 50~to 249                                                                                                              & 1                                                                              & France            & 26           & Male             \\ 
\midrule
\textbf{P12~} & Writer                                                                       & \begin{tabular}[c]{@{}c@{}}Freelance,\\publishers, \\self-publishing \\companies\end{tabular} & N/A                                                                                                                            & 3                                                                              & Canada            & 24           & Male             \\ 
\midrule
\textbf{P13~} & Graphic designer                                                             & Freelance                                                                                     & N/A                                                                                                                            & 2                                                                              & Sri Lanka         & 27           & Male             \\ 
\midrule
\textbf{P14~} & Creative writer                                                              & Freelance                                                                                     & N/A                                                                                                                            & 5                                                                              & South Africa      & 46           & Female           \\ 
\midrule
\textbf{P15~} & Writer                                                                       & Freelance                                                                                     & N/A                                                                                                                            & 6                                                                              & India             & 29           & Female           \\ 
\midrule
\textbf{P16}  & Writer                                                                       & Freelance                                                                                     & N/A                                                                                                                            & 5                                                                              & Germany           & 20           & Female           \\ 
\midrule
\textbf{P17}  & Industrial designer                                                          & Freelance                                                                                     & N/A                                                                                                                            & 7                                                                              & Canada            & N/A          & Female           \\ 
\midrule
\textbf{P18}  & Product designer                                                             & Company                                                                                       &  > 250                                                                                                                           & 1.5                                                                            & Canada            & 23           & Female           \\ 
\midrule
\textbf{P19}  & Software engineer                                                            & Company                                                                                       &  > 250                                                                                                                           &   6                                                                             & Germany           &      32        & Female           
\\ 
\midrule
\textbf{P20}  & Software engineer                                                            & Company                                                                                       &          > 250                                                                                                                   &       9                                                                         & Singapore           &        32      & Female           \\
\bottomrule
\end{tabular}
\label{tab:demographics}
\end{table*}

\subsection{Data Analysis}
The data was analyzed by two authors of this paper. Due to the authors' different but complementary backgrounds for this study (one is an HCI researcher, the other a researcher with a background in law and public policy, having practiced IP law), they analyzed every interview together to ensure they did not miss important details relating to legal or HCI aspects of generative AI. 

They first discussed important things to keep track of relating to AI governance, the 3 Cs, and creative workers' experiences with, and thoughts of generative AI. Afterwards, they conducted an open-ended, iterative thematic analysis of the interviews together, identifying important codes from a few interviews, adjusting codes as they coded more interviews, and then shifted towards a deductive approach once they had a more structured codebook, a common practice in qualitative research~\cite{mcdonald2019reliability}. With this codebook, they formed the codes into relevant themes~\cite{clarke2017thematic}. As the authors analyzed every interview together, which involved discussing areas of disagreement to reach a compromise, there is no interrater reliability score. We identified 121 codes, which fit into 13 different themes. Our codebook can be found in our Supplementary Materials. 

Our \textit{Results} Section focuses on findings that have (to our knowledge) not been discussed in the current literature around generative AI and creative work, such as focusing more on AI governance and regulation of created work. Our codebook contains more detailed codes and themes along with their descriptions, that have not been fully discussed in the paper. 
\section{Results}
In this section, we discuss our findings relating to the 3 Cs (\textit{consent, credit,} and \textit{compensation}) framework, broader generative AI governance findings, and 
disruptions to creative work that cannot be addressed by AI governance.

\subsection{The Role of \textit{Consent, Credit} and \textit{Compensation} in AI Governance}
The 3 Cs of \textit{consent, credit,} and \textit{compensation} have been recommended to mitigate the harms of generative AI on creative workers. This framework suggests that when created output is being used to train AI models, companies need to i) ask creative workers for their consent to use their work, ii) give credit to creative workers whose work has been used to train models, and iii) compensate these creative workers~\cite{3Cs, ccc}.

In our study, however, we found that the role of \textit{consent, credit} and \textit{compensation} was much more nuanced than previous recommendations suggest. We identified the potential benefits to creative workers from this framework, but we also identified areas where this framework may harm workers, as well as identified specifications for how companies can apply this framework responsibly. 

\subsubsection{Consent}
Almost all participants stated they would like to be asked for consent to have models trained on their work. Regarding consent, they would also like to be informed of how their created output would be used, ideally through an employment contract if they were starting a new position, or through email, work meetings, and a mix of formal and informal communication if their work conditions were changing. 

P12 describes the importance of consent: ``\textit{I think transparency is key. I would hate for any author or creative to feel like they were tricked into anything. So I think it goes back to publishers being clear about where work is going and how it's going to be used. If I write something for a company, I want to know it's being published in this country for this long, this is all this information. Or if it is going to feed like feed an AI system, I'd like to know that and then I can make an informed choice.}''

For informed consent, creative workers wanted to be given more information about whether their work would be used to train generative AI models, and if so, how long their work would be retained for (i.e., whether it would be retained after they left the company/publisher/platform). Creative workers stated it was essential to be informed about how their work could be used in a way that is easy to understand, accessible, and with enough time to consider their decision: ``\textit{I would prefer being notified in a way that gives me time to react, voice my own concerns about it. I think if our CEO does it consistently with emails, I think that's fine}'' (P3).

For those who did not feel consent was necessary, it was because they were employees of a company and mentioned that their company owned the work they created during work hours: ``[Worker’s company] \textit{has paid me to work within a particular use case to write code. And so in that case, they own it and it's their code}'' (P19).

\textit{Consent decisions may change with technological advances.}
While consent is important for potentially training AI models on workers' created output, consent decisions change over time as technology advances. Almost all participants said that generative AI is not advanced enough \textit{yet} to replace their own work at this time, but many had concerns about how it could advance in the future, therefore consent, if only asked for once in the beginning, may harm creative workers in the future.

In the words of P8: ``\textit{I'm a bit more comfortable with it (my created output) being used for generating examples because I know the tools aren't that powerful right now. But if in the future the AI tools are powerful enough where it can learn enough from my code to threaten my job prospects elsewhere, then I would I would be a lot more reluctant about giving that permission.}'' A similar sentiment was expressed by P14: ``\textit{Once I've given it (ownership rights) to someone and they've paid me for it, I feel that it's theirs and I don't really mind it being used for training. I'm worried about when we get to the time when it can really start writing a novel that doesn't sound like it's computer generated and what that's going to do to an industry that's already horrifically difficult to get into and to make a living from.}''

\textit{The role of power dynamics in the ability to consent:} Creative workers employed by a company also mentioned the power structures involved in whether they are able to say ``no'' to training generative AI models at work: ``The company owns our a**es, they could do whatever they want with our code, so we can't really tell them they can't use our code for training'' (P10).

This points towards the limits of `consent' in the 3 C's mantra. While being able to object to one's output being used to train AI is important, if work opportunities are hard to come by, and pay is decreasing, the freedom to say ``no'' may exist only in theory; in practice, the worker may be effectively coerced to give permission for their work to be trained on, or face unemployment. This finding is supported by previous work by \citet{chowdhary2023can} which found that workers are unable to provide meaningful consent due to power dynamics, impacting their autonomy and the value of workers' data collected under ``meaningless consent.''


\subsubsection{Credit}
Some participants mentioned that they would like to be attributed if generative AI models were to be trained on their work. They said it was important for them to get recognition, which aligns with research showing that creative workers value getting attributed for their work~\cite{Sprigman, Fisk}. 

However, our findings show that not all creative workers want credit for their work if it were used to train AI models. Of those who did not want to be attributed, there were three reasons:

\textit{Not well-known enough:}
First, they did not consider themselves to be ``famous enough'' to be attributed where it would matter if their name were on the product: ``\textit{I would prefer to be anonymous. I don't think anyone would know who I am anyway. But I do understand that, you know, the bigger, more famous writers would probably definitely want people to know when their work is being used}'' (P14). 

\textit{Companies should get credit:}
Second, they are employees of a company and the company owns their work, as described by P20: ``\textit{All credit should be under the company name, I guess, because we are working for the company.}'' 

\textit{Reducing responsibility:}
Third, some participants explicitly stated they would not like to be attributed because of concerns about what could be created without their knowledge or consent. Some participants did not want to be attributed for their work because they have no control over what users prompt or how their work would be used by generative AI models, and would like to reduce responsibility over AI-generated output. 

Sometimes this concern was related to offensive AI output that could be created from their work: ``\textit{I would be pissed if someone used my work for political reasons because I just wouldn't have any control over it. If someone from a far-right group created images trained on my work I would be really pissed}'' (P11). In light of influential tech companies such as X removing their Ethical AI team with a change in leadership in 2022~\cite{x_AI_ethics}, offensive AI-generated content has been a concern with X's Grok AI image generator~\cite{grok}, and concerns may be heightened for creative workers whose output may be used to train these AI models.

Other times, this concern about receiving credit for training an AI model was about wanting to reduce responsibility over subsequent impacts their work could have: ``\textit{You want to be proud of it (your work), but I know sometimes once it gets out there, you're not happy. You don't know what the outcome is like for a product, you don't know what the outcomes will be like until it’s in somebody else's hands. Sometimes you'd be like, ‘Oh sh**, I didn't realise the impact (my output) would have.’ Like Keurig coffee machines, the inventor of it was, you know, what a convenient and efficient way to have coffee. And then once it was so popularised and he saw how much waste it was producing, he admitted he regretted inventing it. He didn't expect the impact it would have on the environment}'' (P17).

\subsubsection{Compensation}
All participants stated that they would like to be compensated if their work were used to train generative AI models. If creative workers were already working in a company, many said their salary would be adequate, but they would want extra compensation if their work were used in the future, after their employment were to end with that company. 

Creative workers stated that these models could potentially replace their own work, and it is unfair for companies, platforms, and publishers to profit off of this work, especially if they were no longer working with them: ``\textit{I guess a worse scenario is you've sold your work and the publishers profit from it and you're not getting any cut of the share. And they're also doing derivatives of it, and you're not getting anything}'' (P7).

\textit{Acceptance of training generative AI models differs for work vs. personal projects:} In our interviews, creative workers distinguished between generative AI models training on output they create at work, versus personal projects, where compensation may be tricker to implement, and may not fully resolve feelings of unethical use of one's creative work. Creative workers were more likely to accept generative AI models trained on output they produced at work, sometimes seeing it as a part of one's job, and work they are not often very attached to: ``\textit{A lot of the work that I do for products, I personally wouldn't mind (being used for training) specifically because I have a sense of detachment from it in a sense of it's just purely for a product, it's purely for the sellers and my customers}.'' (P18).

Some creative workers mentioned having personal side projects relating to their creative domain, and said they were \textit{not} fine with generative AI models training on their personal projects, as they are more attached to them. As described by P11: ``\textit{If it's just about work, I don't really care, but if it's about my personal projects, I would be mad. The designs I do in my work are for our users, and for the business. But if it's something very personal, I would never want someone to use the stuff that I really care about to make some money, some profit on it.}''

\subsection{Generative AI Governance Principles Beyond the 3 Cs}
Discussions around generative AI governance, especially those relating to creative workers, are in their infancy. However, AI governance is important to address creative workers' needs, represent their voices, and set standards over the usage of generative AI. \textit{Consent, credit,} and \textit{compensation} are important factors to consider when adopting generative AI governance principles, but other AI governance recommendations emerged during our study.


\subsubsection{AI regulation within companies, publishers, and platforms is not yet well-established}
All participants agreed that using generative AI for creative work requires better regulation and governance. Participants suggested that AI governance for addressing creative work should happen at both the governmental (public) level, and the company/platform/publisher (private) level. In the words of P14, ``\textit{I understand that things (technologies) have to move forward, but I do think that governments worldwide have to find something; there are laws that have to be brought into place, and things have to be done.}'' When it comes to regulation at the company/platform/publisher level, P9 said: ``\textit{The companies who make the generative AI models should come with a disclaimer saying `What we use is from existing work made available on the internet that may or may not belong to somebody else.'}''

\textit{Few companies, platforms, and publishers have an AI governance strategy:}
Despite the perceived importance of AI regulation, most participants said the company, publisher, or freelance platform they work with currently has no AI governance strategy around the use of generative AI at work. Oftentimes, creative workers mentioned that their organizations used a reactive mitigation strategy in dealing with generative AI, whereas proactive strategies would be preferred. As described by P12, ``\textit{Recently with this chapter that I've written for a publisher, there was no mention of AI, so I feel like it's either not being considered or the idea is that it's still so nascent and so new that it's not really a threat right now and that it's something that maybe we'll deal with in 5 or 10 years. I think it's not something that the people that I've spoken with are really, like, grappling with in terms of the publishing side.}''

\textit{Widespread AI governance faces challenges:}
There are barriers to the widespread adoption of AI governance, such as the size and financial resources of an organization. In the words of P7, this sometimes leads to the banning of any AI usage to account for the lack of AI governance: ``\textit{I recently submitted to this one magazine, which was very explicit about not allowing any gen AI material. I think the smaller magazines are the ones that tend to be a bit more stringent (around AI usage) partially because they have such limited staffing. Usually they're all volunteer staff and less than 20 people and a lot of them are students or work in academia, so they don't have as much time. So in order to preserve what limited volunteer time that they do have, they just have to have more stringent policies.}''

\subsubsection{Creative workers want a say in how generative AI is used and governed.}
Some creative workers stated that they feel their voices were left out of the generative AI discussions at work, and that it is difficult to communicate their concerns to other communities, especially those building these technologies and those in positions of power to make decisions around AI usage. Very few of our participants were part of unions, which are important for addressing workers’ rights and advocating for workers as a collective. 

In the words of P6, ``\textit{I think my biggest challenge these days as an artist is that artists are typically really bad at thinking in a sort of systematic way about their own work and on the whole, we're a very sort of intuitive bunch. So when I see artists try to make a case to others who are not artists about why their work is important, how do you get (convey) that information to someone else?}'' This challenge was not expressed by participants from all creative domains. However, previous research has found that creative professionals in other domains, such as programmers, are also limited in the amount of power they have to express their ethical concerns~\cite{widder2023s}.

\subsubsection{Generative AI models being trained on copyrighted content was a common concern}
Many creative workers in our sample had concerns about copyrighted content and data quality, not only about generative AI models potentially training on their own work but also about how they could potentially break copyright laws if they were using AI-generated work. 

Because of these concerns, some participants were advised by their companies to avoid using generative AI for sensitive tasks, such as production code or final design versions, but were told it was fine to use it for drafts. Creative workers, particularly those in design and visual art, stated that they preferred to use generative AI tools that were trained on licensed or public domain images, such as Adobe Firefly~\cite{firefly}, to avoid potential legal and ethical issues. 

\subsubsection{Creative workers want more regulation to focus on differentiating between AI-made and human-made work}
Creative workers stated that a threat to their current work is the lack of differentiation being made between human-made and AI-made work, which is where AI governance can be helpful.
In the words of P5: ``\textit{There needs to be a proper regulation of sorts on how to use and how to label work that is made by AI versus work that's truly made by humans.}'' Better labeling of what is and is not AI-generated, such as through a filter on platforms or having badges indicating who/what created output, was deemed an important step towards making generative AI more responsible. 

Creative workers in design may rely on other tools, such as image or code libraries, to gather inspiration or solve a problem, but AI-generated content is also spreading to these libraries, causing concerns of not only copyright but also decreasing the quality of these platforms. ``\textit{As a designer, I use a few platforms. I take some images there, I pay a subscription fee and those images help me in my creative process. If I go on those platforms and I'm looking for an image, there's an abundance of AI artwork trying to pass as not-AI. That is very difficult for me to work with because the platform is not keeping a tight grasp on it and it makes my process of searching for inspiration time-consuming and difficult}'' (P9).

\subsection{Changes to Creative Work Not Addressed by Generative AI Governance}
In this section, we discuss changes and disruptions to creative work that cannot be addressed by (generative) AI governance, which may require creative workers to adapt to the changing working environment on their own. 

\subsubsection{How creative workers adapt to generative AI}
While creative workers expressed excitement, worries, and doubts about generative AI, participants said that AI is likely here to stay and they will need to adapt to be competitive in their field. This is particularly important for protecting one's career in case AI governance does not adequately address the interests of creative workers.

\textit{Better prompting:}
Creative workers said it is important to become better at prompting to get more precise output:
``\textit{Everyone can generate an image using Midjourney, but it takes someone that really understands how to use the perfect words and create a prompt such that you generate the perfect picture you need. So that's a skill you can market or you can sell}'' (P13).

\textit{Augmenting skills:}
Creative workers said generative AI should be used to augment skills rather than replace them because this will improve work quality while ensuring output maintains a level of human thoughtfulness that AI typically lacks. As described by P6: ``\textit{I think a good example was what I think were generative AI tools on the new Spider-Verse movie. It's the style that they got, which is that sort of very hand-drawn style, and they use generative AI to make sure that the lines looked a certain way across frames. It would have been impractical to do if you had had to rely on humans and so they did it with AI, and it allowed them to get a really great look. And it was a human being saying, `Here is the style, here is the target, and let's get the thing (AI) to do the grunt work in the back end.’ That's labor saving}.''

\textit{Accessibility:}
Creative workers said generative AI can improve accessibility for learning new skills faster. P8 said: ``\textit{On the hopeful side, I think that (generative AI) might lead to a lot better involvement in, say like more complicated open source topics or projects that require a lot more specialised skills. So, for example, I think a few years ago, there was some concern that the maintainers of OpenSSL were only a handful of people across the world, and for good reason because you need to be very skilled with cryptography to work with OpenSSL. Maybe in the future some generative AI tool would make it a lot easier to get into those advanced cryptography topics and be able to explain some of the more esoteric code in that code base and then maybe that would lead to a wider group of people better qualified and able to contribute good patches to these projects.}''

\subsubsection{Generative AI may disrupt creative industries}
Creative workers stressed that generative AI, when used in a way that replaces workers rather than aiding them (such as helping with menial tasks or augmenting their own skills), can negatively impact the quality of output. 

\textit{Loss of respect for creative work:}
Creative workers feared that there would be a loss of respect for the craft, and making non-experts believe anyone can become a writer, designer, artist, or programmer without years of training and practice,  P16 ``\textit{I've realized that many people, they claim to be writers on Fiverr, but all they do is just use AI.}''

\textit{Loss of work for creative workers:}
Connected to a loss of respect for creative work, the rise in generative AI's popularity has impacted some participants’ livelihoods, especially those working as freelancers: ``\textit{When it comes to writing, I realised people order way less from me now since AI is a thing now. A couple of years ago, I had many orders, and many people wanted to work with me}'' (P16). This finding is supported by previous work showing that online freelancers in writing and code-related jobs prone to automation are indeed losing work~\cite{demirci2023ai}, and that certain jobs, such as writing and programming, are more prone to being replaced by AI~\cite{felten2021occupational, demirci2023ai}. 

\textit{Lack of human thoughtfulness:} 
Participants stated that AI-generated content does not have the level of human thoughtfulness that human-created output has: ``\textit{I'm really scared that people think they can be designers just because they can make some graphics with AI. Design is really more important and much deeper than that. It is not only about how things look, but also the experience and the problem that it solves}'' (P11).

\textit{Shallow understanding of creativity:} 
Creative workers stressed that generative AI can lead to a shallower understanding of the creative process. In the words of P3: ``\textit{Instead of doing the learning themselves, which, you know, they might find is tedious, they now just delegate it to ChatGPT and all they're doing is kind of like pasting code around.}'' Previous work by ~\citet{messeri2024artificial} has shown that AI can lead to an ``illusion of understanding'' where users may think they know more about a subject, and therefore produce more work, while only having a shallow understanding of concepts.

\textit{Elimination of creativity:} 
Creative workers also commonly mentioned that generative AI can eliminate creativity that naturally comes with human-made work because generative AI is good at mimicking, but not necessarily creating something original. In the words of P6: ``\textit{As far as what I'm seeing, I think it's going to be an extension of trends that already existed and it's going to make it much worse, which is everything looking the same. It's sequels after sequels after sequels because it changes the incentives. It's already difficult to get anybody to sign onto anything new or to spend any amount of money on anything new. And, you'll get, you know, \textit{Ghostbusters 17}... Isn't it better to make new things and add to culture and make today different than what yesterday was like?}''
\section{Discussion}
Our interview study with 20 creative workers in visual art and design, programming, and writing investigated what creative workers want out of generative AI governance, focusing on the 3 Cs (\textit{consent, credit,} and \textit{compensation}), and generative AI governance recommendations beyond the 3 Cs. Our findings confirm previous research showing that creative professionals think generative AI, in its current state, still requires human input and is therefore not a very serious concern for job replacement yet~\cite{li2024user, li2024value, palani2024evolving}. Instead, many creative workers view generative AI simply as a tool, echoing a finding from \citet{10.1145/3613904.3642133}. When creative workers in our study did say they used generative AI, many had copyright infringement concerns about the data generative AI models are trained on, showing that there is an increasing demand for copyright-free content, a finding which also has been surfaced in previous work~\cite{han2024understanding}. 
 
Understanding creatives' concerns about work replacement and the role generative AI plays helps us better understand how it can be harmful to creatives and how it can be adapted to help creatives succeed. However, it is also important to focus on generative AI governance and workers' protection policies to address workers' needs. We expand upon previous work about the impacts of generative AI on creative work(ers) by investigating how to address workers' needs better through creative worker-centric generative AI governance. Specifically, we focus on the 3 Cs framework of \textit{consent, credit,} and \textit{compensation}, which lacks empirical research on whether creative workers find this framework useful as a protection mechanism from the potential harms generative AI may bring.

We found that generative AI governance is perceived as important to protect creative workers' rights and livelihoods from the potential harms of generative AI, but there is a gap between what workers want, and what is currently in place, not only from governments, but also from companies, publishers, and freelance platforms that creative workers commonly work with. We also found that the 3 Cs play a more nuanced role than current discourse may suggest~\cite{3Cs, ccc}, with both the potential to protect and harm creative workers, therefore requiring careful consideration if regulators and other stakeholders were to implement this framework into practice. 

\subsection{Acceptable and Unacceptable Uses of Generative AI}
Creative workers shared mixed feelings towards generative AI. When it is used in an acceptable manner, it can result in an effective collaboration between creative workers and AI, but when used in an unacceptable, unethical manner, it can harm creative workers. Therefore, we recommend that more consideration be placed into how generative AI is trained, used and governed in relation to creative work to enhance the ethics of these models.

In the EU context, the AI Act considers employment and labor markets as high-risk domains for the application of AI systems generally (see Annex III, paragraph 4), but it mostly focuses on the application of the AI systems in this domain, e.g., the use of AI systems in recruitment, work organization, or evaluation of workers. However, the \textit{creation} of AI systems (including the data and data pipelines used to train the systems) can also create risks in the employment context and labor markets, as we discussed in this paper. From a policy perspective, it may become relevant to consider regulating the data, data pipelines, and organizational settings in which AI systems are made and trained on (including the platforms and employment contexts in which our interviewees worked), in addition to their outputs and impacts.

In the EU AI Act, it is specifically mentioned that AI systems should be classified as ``high risk'' when they ``may have an appreciable impact on future career prospects, livelihoods of those persons and workers’ rights'' (recital 57). As such, we recommend that AI companies, employers, freelancing platforms, and publishers abide by these regulations and not engage in uses of AI which may impact the rights or careers of creative workers which may be deemed as an unacceptable use of AI.



\subsection{Power Dynamics in the Generative AI and Creative Work Landscape}
In our study, we identified some underlying issues of power dynamics which exist in the generative AI and creative work landscape, evidenced by three points:

\paragraph{Unpredictable power shifts with emerging technologies.} Creative workers in our interviews expressed worries about losing control over their work, such as it being used for training without their knowledge or consent. Creatives often give up rights to their work when signing contracts with companies and clients, but it is impossible to consent to licensing, etc., of rights for future unspecified uses and unknown technologies.

\paragraph{Information asymmetries.} As the inputs used to train generative AI models are largely unknown, there are information asymmetries between creative workers whose work may be used to train generative AI models, and companies building these models. Thus, creative workers in our study expressed some doubts for the future of creative work, and wanted more transparency from employers, publishers, and platforms they were working with. This can be in the form of making generative AI models more transparent with regard to how they are trained, and better informing workers about if and when their work will be used to train such models, along with proper consent, credit, and compensation considerations.

\paragraph{Lack of established AI governance and regulation.} Our study found that most organizations which creative workers worked at, either as an employee or freelancer, lacked an established AI governance policies. Sometimes, this led to reactive, rather than proactive approaches to AI governance, or ignoring AI (such as banning its usage) altogether. This lack of lower-level AI governance, supplemented with a lack of widely adopted AI regulations around the world, can also lead to a power imbalance between creative workers and companies potentially training on their work.

\subsection{Implications for Workers} 
\textit{Generative AI impacts creative workers in companies vs. freelancing differently.}
Our findings indicate that different forms of employment, and therefore creative domain, are impacted differently by generative AI. Generally, freelancers in our participant sample had more concerns about generative AI replacing their jobs because they tend to have less stable income than company employees. This is reflected in previous work, which shows that freelancers try to find work that does not rely on platforms or clients to build their credibility and find stability~\cite{blaising2022managing}. 

Additionally, we found that creative workers employed by a company tended to be less attached to their work compared to freelancers. As such, these participants were generally fine with companies training AI models on their work, but only while they are employed by them. While employees of a company are more secure in their employment compared to freelancers, they can also experience harm if their work is used to train models when they are no longer employed by a company, especially without their consent or further compensation.

Within our participant sample, all programmers we interviewed were employed by a company, whereas writers, designers, and artists tended to be a mix of freelancers and employees. We observed differences in how secure creative workers in different domains felt with regard to generative AI. While many participants expressed some level of concern about (future) job replacement by AI, the level of immediate threat differed. Some writers, designers, and artists we interviewed, particularly those who are freelancing, have seen a decrease in work due to generative AI. In contrast, programmers also expressed feeling worried, but were not as impacted by generative AI yet, as they have more present-day job stability.

\textit{Collective action can help protect creative workers.}
Due to the different forms of work creative workers may be involved in, ranging from being an employee of a company to freelancing on a platform or freelancing with publishers, we need generative AI governance adapted to different working conditions, with special policies to protect more precarious working situations such as freelancing. 

Echoing previous findings~\cite{game_conf}, most participants in our sample were not part of a union, and freelancing makes unionizing difficult. Formally, most freelancers cannot really `unionize' (in the sense of joining an association legally recognized as a `union') and cannot do collective bargaining within the framework of most labor laws (for example, the US National Labor Relations Act), as this is a violation of competition law. However, collective action efforts can strengthen workers' voices and better protect them~\cite{widder2023s}. As creative workers are in a vulnerable position and generative AI is impacting their careers, workers need to be represented in generative AI discussions, both at the local company/publisher/freelance platform level, and at broader governmental levels. 

\subsection{Implications for Companies, Publishers, and Platforms}
Companies, publishers, and platforms are potential avenues that creative workers may work at and can act as an intermediary between creative workers and the ``outside world,'' such as the public, clients, or other entities. Therefore, companies, publishers, and platforms can play a role in protecting the workers they work with. 

\textit{Workers want better-established generative AI governance strategies.}
Many creative workers in our study wanted the companies, publishers, and platforms they worked with to adopt a more proactive and thorough generative AI governance strategy to address their needs. Specifically, this generative AI governance strategy could include more transparency around AI usage, such as clarifying when generative AI can or cannot be used, the conditions around its usage, being notified if/when their work may be used to train AI models, and wanted more done to differentiate between AI- and human-made work. 

Creative workers who worked on freelance platforms also wanted better filters to differentiate between AI-generated and human-generated work. Freelance platforms should have more rules around who can work on their platforms as a ``freelancer,'' since our study found that some freelancers complained about other freelancers using AI inappropriately in their services. Importantly, companies, publishers, and platforms should also gain consent from creatives around generative AI usage, indicating whether their work could be potentially used to train generative AI models, in the present or in the future.

Additionally, we found that creative workers working for differently-sized companies and publishers had different levels of generative AI adoption and rules at the local level. Those who worked with bigger, therefore usually better resourced, companies and publishers often reported that generative AI was more likely to be used at work. As such, these better-resourced companies and publishers sometimes had some set of formal or informal rules around the use of generative AI compared to those with fewer resources.

\subsection{Implications for Regulation and Policy}
The law can protect a weaker party from a more powerful party who may misuse their power or knowledge. Often, copyright, IP, and patent law can be weaponized and used against creators, especially by ``bigger players'' with more resources and legal power to protect their assets~\cite{doctorow2023internet}. There are some ways to give power back to other stakeholders and promote rights for creative workers, and part of this is through effective AI regulation and policy, especially those aimed at giving weaker parties more power. 

More effective generative AI governance that considers creative workers' needs and preferences, more creative work protections are possible. Generative AI regulation, at both governmental and local (company/platform/publisher) levels, should focus on improving creative workers’ protections and involving workers’ voices. In our discussion, we present how, through the lens of AI governance, we can re-distribute the power that big tech corporations have over the other affected stakeholders.

\textit{Regulation vs. Innovation.} One question that arises in considering potential new policies addressing generative AI governance and in potential updates to existing regulatory frameworks such as intellectual property or data protection law, is whether such regulatory developments could stifle or slow innovation, and potentially thereby hamper economic growth and reduce consumer welfare. In this context specifically, the question is whether such regulatory developments would, or could, slow the development of generative AI technologies. Regulations may address creative workers’ livelihoods and rights, at least in the short term; however, it may also slow adoption and in-country development of generative AI technology, and thereby potentially cause long-term economic harm (e.g., through loss of economic growth opportunities, competitiveness, or indigenous technological capability).

There may not always be a tradeoff between regulation and innovation. Anu Bradford argues that there are features in the US legal and tech ecosystem that the EU lacks, rather than the EU's focus on regulating technologies, that have allowed for US-based tech companies to gain more international distinction~\cite{bradford2024false}. These differences include the EU consisting of multiple nations, languages, and local regulations making it difficult to scale and grow their businesses, less funding for EU companies and their innovations, the EU having a culture of being averse to risk and costly bankruptcy laws making risks extra risky, and the US tech sector's ability to attract global tech talent~\cite{bradford2024false}.

Additionally, there are different types of innovation, and not all of them contribute equally to economic growth or consumer/social welfare. Especially in the context of increasing climate change impacts and political polarization, policymakers may feel a need to assess the potential economic and social benefits---and risks---of new technologies with frameworks beyond the traditional economic assumptions that technological innovation leads to economic growth, which leads in turn to broad-based improvements to consumer welfare~\cite{stiglitz2009report, raworth2018doughnut, mazzucato2021mission, o2018eu, macron2023green}. The net economic and social benefit of generative AI systems, and therefore the potential costs of establishing regulatory frameworks that might slow their development or adoption, is in this context extremely difficult to estimate.

\textit{Recommendations for applying the 3 Cs into regulation and practice.}
Based on our findings, we present implications for regulation and policy: 

\paragraph{Consent:} 
Consent is necessary for AI systems to be responsible~\cite{singh2022consent}. As such, policymakers could require employers and other contracting counterparties (e.g., publishers, platforms) to explicitly inform employees and other creative workers if they intend to use creative workers’ work to train generative AI systems, or allow other parties (e.g., firms building and operating such systems) to do so. There could also be efforts made to establish requirements for valid consent (e.g., Art. 7 GDPR); namely, that consent must be \textit{informed} and \textit{genuine} (i.e., truly freely given); i.e., workers must be effectively protected from any direct or indirect penalty, punishment, or disadvantage if they decline to give consent.

In our study, many creative workers mentioned that they were fine with generative AI models being trained on their work, but had concerns about how generative AI may advance and potentially replace them in the future. As such, we recommend that consent should not be a one and done situation; creative workers should be asked for explicit and informed consent when they are first employed with a company or work with a publisher/freelance platform, in addition to being asked for consent from companies building and operating generative AI systems.

However, consent should not stop there. Asking for explicit and informed consent has to \textit{progress} with advances in generative AI technologies and updates to working conditions. Workers should be given adequate information to make an informed decision, and workers should be given enough time and notice to make this consent decision (such as why, how, and how long their work would be used), along with an alternative if workers decide to withdraw consent. Previous work by ~\citet{zou2018ve} has shown that most users will not pay attention to important security-related notifications, so it is important that these consent updates be delivered in various forms that workers find out about work updates (e.g., directly from their supervisor, during group meetings, email, etc.), and they should be quick and easy to understand.

Given the potentially significant limitations to the effectiveness of individual consent for governing generative AI, establishing collective or representative consent mechanisms (e.g., ‘works councils’ in some European jurisdictions) could be a consideration. Refusal to provide consent after an advance in technologies should not invalidate or otherwise impact previously given consent, e.g., by creating an obligation to ‘retroactively’ remove inputs to already-trained systems; cf. e.g. Art. 7(3) GDPR, Sentence 2.)

\paragraph{Credit:}
According to the 3 Cs framework, crediting creative workers is important~\cite{ccc, 3Cs}, but whether creative workers \textit{want} credit for training generative AI models is more nuanced in practice. Sometimes, creative workers are employed by a company that already owns the rights to the work they create during work time, leading some creatives to think their company should get credit. Other times, creative workers have concerns about their reputation and the output that could potentially come out of models trained on their work, and credit may do more harm than good.

In theory, these nuances of credit are captured and protected through moral rights enshrined in Article 6bis of the Berne Convention~\cite{6bisBerne}. In the EU, moral rights are born out of creative workers’ inalienable personality~\cite{rigamonti} and, therefore ‘independently of the author’s economic rights, and even after the transfer of the said rights, the author shall have the right to claim authorship of the work and to object to any distortion, mutilation or other modification of, or other derogatory action in relation to, the said work, which would be prejudicial to his honor or reputation’~\cite{6bisBerne}. 

Moral rights within the EU and worldwide are not harmonized or interpreted in the same manner~\cite{rigamonti}. With some variations, the French intellectual property code (Code de la propriété intellectuelle) and German law on copyrights and related laws (Gesetz über Urheberrecht und verwandte Schutzrechte) define moral rights~\cite{cotter} to include:

\begin{itemize}
    \item Right to disclosure: right to determine when the work is ready to be disclosed to the public;
    \item Right to correct or withdraw works that were previously disclosed to the public;
    \item Right to attribution;
    \begin{itemize}
        \item Right to misattribution: being attributed as the author of another’s work or having another attributed for the author’s own work;
        \item Right against non-attribution: omission of one’s name from one’s work;
        \item Right to publish anonymously or pseudonymously;
        \item Right to void a promise to publish anonymously or pseudonymously; and
    \end{itemize}
    \item Right to respect of the work/ right to integrity.
\end{itemize}


In the EU, authors of databases are granted moral rights per the member state’s respective legislation~\cite{databasedir}. Given that model development is a multi-stakeholder endeavor~\cite{yurrita} and that authors of databases have moral rights, how the right to attribution would trickle down through the model development pipeline, especially if synthetic data is used for training, would be another source of a legal difficulty. ~\citet{miernicki} argued that moral rights are not just about the personality in the creative workers’ work but also the creativity and originality that the creative worker’s personality brings to their work.

For foolproof long-term solutions, we second the idea brought forth by ~\citet{miernicki} to urgently reframe and ideologically decide on legal frameworks for ownership, originality, and creativity of generative AI models and their outputs that can address stakeholder needs better while incentivizing creation of original content creation.

\paragraph{Compensation:}
Our study found that all creative workers in our sample wanted to be compensated if their work was used to train generative AI models, either in the form of employment with a company, or additional compensation if one were a freelancer or no longer employed by that company. For work projects, creative workers tended to be less attached, therefore were fine with their workplace compensating them through employment, and paying them after they finish their role. For personal projects, compensation may not be enough, and it is imperative that the informed consent of the original creator(s), in addition to establishing ownership structures and compensation, are additional considerations.

With regard to compensation, policymakers could establish or encourage professional associations or trade unions to establish normative baselines for compensation (e.g., royalties) for creative workers whose work is used (or anticipated to be used) to train generative AI systems. Such ‘norms’ could take the form of model contracts applicable to different contractual relationships (e.g., employment, freelance) and situations (e.g., use of the generative AI system to produce derivative works for commercial use after termination of the employment or freelance contract).

Data dividends have been suggested in previous research to more fairly distribute the gains of AI across different stakeholders involved in training and creating AI~\cite{vincent2024data}. To calculate how much to compensate creative workers, a data value determination may be conducted, which would assess the benefits, costs, and quality in determining the value of creative outputs~\cite{bodendorf2024business}. Value determination adds additional nuance to the 3 Cs framework, as there are currently many challenges and methods to determine the value of data~\cite{fleckenstein2023data, coyle2024value}, and users' perceptions of what data is valuable differs from how companies view user data~\cite{kyi2024doesn}.

\subsection{Limitations}
We recognize our study has limitations. First, we only interviewed creative workers from three domains relevant to the HCI and computer science community (visual art/design, writing, and programming). However, we recognize that other creative domains have been impacted by generative AI. Therefore, we cannot be certain that our results are generalizable to all domains such as music, acting, content creation, etc. 

Second, there could be participant self-selection in our sample, where those who feel more opinionated about generative AI, or those who are in more secure positions to speak up, may be more willing to be interviewed about it. We also may have missed participants who are under non-disclosure agreements (NDAs) and therefore limited in what they can share about their work(place). 

Third, as we used a snowballing technique to recruit participants, there was less diversity in terms of age, experience, region, and perhaps opinions, therefore they may not fully represent the creative community. We tried mitigating these effects by interviewing participants we did not know through Fiverr.

\subsection{Future Directions}
In this paper, we investigated how creative workers would like AI governance to better address their needs, focusing on the role of \textit{consent, credit,} and \textit{compensation}, in addition to additional AI governance considerations. In the future, more work could further explore the 3 Cs and how to apply them concretely in practice. Additionally, as our study did not go in-depth for differences between different creative professions, we recommend follow-up work that deeply investigates how different professions have been impacted by generative AI. 

For \textit{consent,} further investigation is needed to understand how to design consent mechanisms for creative workers. There are lessons to be learned from previous applications of consent such as consent notices under the EU's General Data Protection Regulation (GDPR), consent in medical contexts, or for human subjects research to understand where these consent applications went wrong, or what they are doing correctly to collect informed consent from users.

For \textit{credit,} more work can be done to understand how and when to give creative workers credit, and further investigate ownership structures and moral rights in the generative AI pipeline. 

Lastly, for \textit{compensation,} research can be done to better understand how much is reasonable to compensate creative workers such as understanding workers' valuation of their created output, and how to set up compensation structures, such as through data dividends~\cite{vincent2024data}.
\section{Conclusion}
In our study, we interviewed 20 creative professionals working across visual art and design, writing, and programming to better understand how creative output can be better governed in the age of generative AI. We provide a more nuanced discussion of \textit{consent, credit,} and \textit{compensation}, identifying how these concepts may benefit or harm creative workers, and how to apply them responsibly in practice. Our study also identified other areas where generative AI governance can be improved to better protect creative workers beyond the 3 Cs, and disruptions to creative work that generative AI governance may not be able to address. 

Currently, the future of generative AI may appear to be in the hands of a few well-resourced large technology companies, with implications on affected stakeholders. In our paper, we discuss how effective generative AI governance and regulation could help redistribute power to different stakeholders, such as creative workers, and how regulation and policy can incorporate worker-centered generative AI governance.

\
\begin{acks}
We would like to thank the anonymous CHI reviewers and our participants for taking the time to share their insights and experiences for this study. We also thank Yixin Zou and Selin Fidan for the helpful discussions and providing feedback on our paper draft. Time for coauthor Six Silberman was funded partially by the European Research Council under the European Union's Horizon 2020 research and innovation program (grant agreement no. 947806).
\end{acks}



\bibliographystyle{ACM-Reference-Format}
\bibliography{sample-base}


\begin{thebibliography}{92}


\ifx \showCODEN    \undefined \def \showCODEN     #1{\unskip}     \fi
\ifx \showDOI      \undefined \def \showDOI       #1{#1}\fi
\ifx \showISBNx    \undefined \def \showISBNx     #1{\unskip}     \fi
\ifx \showISBNxiii \undefined \def \showISBNxiii  #1{\unskip}     \fi
\ifx \showISSN     \undefined \def \showISSN      #1{\unskip}     \fi
\ifx \showLCCN     \undefined \def \showLCCN      #1{\unskip}     \fi
\ifx \shownote     \undefined \def \shownote      #1{#1}          \fi
\ifx \showarticletitle \undefined \def \showarticletitle #1{#1}   \fi
\ifx \showURL      \undefined \def \showURL       {\relax}        \fi
\providecommand\bibfield[2]{#2}
\providecommand\bibinfo[2]{#2}
\providecommand\natexlab[1]{#1}
\providecommand\showeprint[2][]{arXiv:#2}

\bibitem[Adobe(2024)]%
        {firefly}
\bibfield{author}{\bibinfo{person}{Adobe}.} \bibinfo{year}{2024}\natexlab{}.
\newblock \bibinfo{title}{{Adobe Firefly vs. DALL·E 3: Express your vision with the right AI art generator for you}}.
\newblock \bibinfo{howpublished}{\url{https://www.adobe.com/products/firefly/discover/firefly-vs-dalle.html}}.
\newblock


\bibitem[AI(2024)]%
        {public_data_openAI}
\bibfield{author}{\bibinfo{person}{Open AI}.} \bibinfo{year}{2024}\natexlab{}.
\newblock \bibinfo{title}{{How ChatGPT and our language models are developed}}.
\newblock \bibinfo{howpublished}{\url{https://help.openai.com/en/articles/7842364-how-chatgpt-and-our-language-models-are-developed}}.
\newblock


\bibitem[Alba and Chang(2024)]%
        {youtube_openAI}
\bibfield{author}{\bibinfo{person}{Davey Alba} {and} \bibinfo{person}{Emily Chang}.} \bibinfo{year}{2024}\natexlab{}.
\newblock \bibinfo{title}{{YouTube Says OpenAI Training Sora With Its Videos Would Break Rules}}.
\newblock \bibinfo{howpublished}{\url{https://www.bloomberg.com/news/articles/2024-04-04/youtube-says-openai-training-sora-with-its-videos-would-break-the-rules?sref=10lNAhZ9&embedded-checkout=true}}.
\newblock


\bibitem[Allyn(2023)]%
        {NYT_openai}
\bibfield{author}{\bibinfo{person}{Bobby Allyn}.} \bibinfo{year}{2023}\natexlab{}.
\newblock \bibinfo{title}{{'New York Times' sues ChatGPT creator OpenAI, Microsoft, for copyright infringement}}.
\newblock \bibinfo{howpublished}{\url{https://www.npr.org/2023/12/27/1221821750/new-york-times-sues-chatgpt-openai-microsoft-for-copyright-infringement}}.
\newblock
\newblock
\shownote{[Online; accessed 07-March-2024]}.


\bibitem[Andersen et~al\mbox{.}(2023)]%
        {Sarah}
\bibfield{author}{\bibinfo{person}{Sarah Andersen}, \bibinfo{person}{Ors v Stability~AI}, {and} \bibinfo{person}{Ors}.} \bibinfo{year}{2023}\natexlab{}.
\newblock \bibinfo{howpublished}{\url{https://fingfx.thomsonreuters.com/gfx/legaldocs/myvmogjdxvr/IP\%20AI\%20COPYRIGHT\%20complaint.pdf}}.
\newblock


\bibitem[Baria and Cross(2021)]%
        {baria2021brain}
\bibfield{author}{\bibinfo{person}{Alexis~T Baria} {and} \bibinfo{person}{Keith Cross}.} \bibinfo{year}{2021}\natexlab{}.
\newblock \showarticletitle{The brain is a computer is a brain: neuroscience's internal debate and the social significance of the Computational Metaphor}.
\newblock \bibinfo{journal}{\emph{arXiv preprint arXiv:2107.14042}} (\bibinfo{year}{2021}).
\newblock


\bibitem[Blaising and Dabbish(2022)]%
        {blaising2022managing}
\bibfield{author}{\bibinfo{person}{Allie Blaising} {and} \bibinfo{person}{Laura Dabbish}.} \bibinfo{year}{2022}\natexlab{}.
\newblock \showarticletitle{Managing the transition to online freelance platforms: Self-directed socialization}.
\newblock \bibinfo{journal}{\emph{Proceedings of the ACM on Human-Computer Interaction}} \bibinfo{volume}{6}, \bibinfo{number}{CSCW2} (\bibinfo{year}{2022}), \bibinfo{pages}{1--26}.
\newblock


\bibitem[Blog(2024)]%
        {copilot}
\bibfield{author}{\bibinfo{person}{Stack~Overflow Blog}.} \bibinfo{year}{2024}\natexlab{}.
\newblock \bibinfo{title}{{Is AI making your code worse?}}
\newblock \bibinfo{howpublished}{\url{https://stackoverflow.blog/2024/03/22/is-ai-making-your-code-worse}}.
\newblock


\bibitem[Bodendorf and Franke(2024)]%
        {bodendorf2024business}
\bibfield{author}{\bibinfo{person}{Frank Bodendorf} {and} \bibinfo{person}{J{\"o}rg Franke}.} \bibinfo{year}{2024}\natexlab{}.
\newblock \showarticletitle{What is the business value of your data? A multi-perspective empirical study on monetary valuation factors and methods for data governance}.
\newblock \bibinfo{journal}{\emph{Data \& Knowledge Engineering}}  \bibinfo{volume}{149} (\bibinfo{year}{2024}), \bibinfo{pages}{102242}.
\newblock


\bibitem[Boucher et~al\mbox{.}(2024)]%
        {10.1145/3613904.3641889}
\bibfield{author}{\bibinfo{person}{Josiah~D Boucher}, \bibinfo{person}{Gillian Smith}, {and} \bibinfo{person}{Yunus~Do\u{g}an Telliel}.} \bibinfo{year}{2024}\natexlab{}.
\newblock \showarticletitle{Is Resistance Futile?: Early Career Game Developers, Generative AI, and Ethical Skepticism}. In \bibinfo{booktitle}{\emph{Proceedings of the CHI Conference on Human Factors in Computing Systems}} (<conf-loc>, <city>Honolulu</city>, <state>HI</state>, <country>USA</country>, </conf-loc>) \emph{(\bibinfo{series}{CHI '24})}. \bibinfo{publisher}{Association for Computing Machinery}, \bibinfo{address}{New York, NY, USA}, Article \bibinfo{articleno}{173}, \bibinfo{numpages}{13}~pages.
\newblock
\showISBNx{9798400703300}
\urldef\tempurl%
\url{https://doi.org/10.1145/3613904.3641889}
\showDOI{\tempurl}


\bibitem[Bradford(2024)]%
        {bradford2024false}
\bibfield{author}{\bibinfo{person}{Anu Bradford}.} \bibinfo{year}{2024}\natexlab{}.
\newblock \showarticletitle{The False Choice Between Digital Regulation and Innovation}.
\newblock \bibinfo{journal}{\emph{Northwestern University Law Review}} \bibinfo{volume}{118}, \bibinfo{number}{2} (\bibinfo{year}{2024}).
\newblock


\bibitem[Butcher and Beridze(2019)]%
        {butcher2019state}
\bibfield{author}{\bibinfo{person}{James Butcher} {and} \bibinfo{person}{Irakli Beridze}.} \bibinfo{year}{2019}\natexlab{}.
\newblock \showarticletitle{What is the state of artificial intelligence governance globally?}
\newblock \bibinfo{journal}{\emph{The RUSI Journal}} \bibinfo{volume}{164}, \bibinfo{number}{5-6} (\bibinfo{year}{2019}), \bibinfo{pages}{88--96}.
\newblock


\bibitem[Chowdhary et~al\mbox{.}(2023)]%
        {chowdhary2023can}
\bibfield{author}{\bibinfo{person}{Shreya Chowdhary}, \bibinfo{person}{Anna Kawakami}, \bibinfo{person}{Mary~L Gray}, \bibinfo{person}{Jina Suh}, \bibinfo{person}{Alexandra Olteanu}, {and} \bibinfo{person}{Koustuv Saha}.} \bibinfo{year}{2023}\natexlab{}.
\newblock \showarticletitle{Can workers meaningfully consent to workplace wellbeing technologies?}. In \bibinfo{booktitle}{\emph{Proceedings of the 2023 ACM Conference on Fairness, Accountability, and Transparency}}. \bibinfo{pages}{569--582}.
\newblock


\bibitem[Cihon et~al\mbox{.}(2020)]%
        {cihon2020should}
\bibfield{author}{\bibinfo{person}{Peter Cihon}, \bibinfo{person}{Matthijs~M Maas}, {and} \bibinfo{person}{Luke Kemp}.} \bibinfo{year}{2020}\natexlab{}.
\newblock \showarticletitle{Should artificial intelligence governance be centralised? Design lessons from history}. In \bibinfo{booktitle}{\emph{Proceedings of the AAAI/ACM Conference on AI, Ethics, and Society}}. \bibinfo{pages}{228--234}.
\newblock


\bibitem[Clarke and Braun(2017)]%
        {clarke2017thematic}
\bibfield{author}{\bibinfo{person}{Victoria Clarke} {and} \bibinfo{person}{Virginia Braun}.} \bibinfo{year}{2017}\natexlab{}.
\newblock \showarticletitle{Thematic analysis}.
\newblock \bibinfo{journal}{\emph{The journal of positive psychology}} \bibinfo{volume}{12}, \bibinfo{number}{3} (\bibinfo{year}{2017}), \bibinfo{pages}{297--298}.
\newblock


\bibitem[Commission(2024)]%
        {ai_act}
\bibfield{author}{\bibinfo{person}{European Commission}.} \bibinfo{year}{2024}\natexlab{}.
\newblock \bibinfo{title}{{AI Act}}.
\newblock \bibinfo{howpublished}{\url{https://digital-strategy.ec.europa.eu/en/policies/regulatory-framework-ai}}.
\newblock


\bibitem[Concord Music~Group and v~Anthropic~PBC(2023)]%
        {Concord}
\bibfield{author}{\bibinfo{person}{INC Concord Music~Group} {and} \bibinfo{person}{Ors v Anthropic~PBC}.} \bibinfo{year}{2023}\natexlab{}.
\newblock \bibinfo{howpublished}{\url{https://storage.courtlistener.com/recap/gov.uscourts.tnmd.96652/gov.uscourts.tnmd.96652.1.0.pdf}}.
\newblock


\bibitem[Constitution({[n.\,d.]})]%
        {USConstitution}
\bibfield{author}{\bibinfo{person}{US Constitution}.} \bibinfo{year}{[n.\,d.]}\natexlab{}.
\newblock \bibinfo{howpublished}{Art. I, § 8(8)}.
\newblock
\urldef\tempurl%
\url{https://constitution.congress.gov/browse/essay/artI-S8-C8-1/ALDE_00013060/}
\showURL{%
\tempurl}


\bibitem[Cotter(1997)]%
        {cotter}
\bibfield{author}{\bibinfo{person}{Thomas~F Cotter}.} \bibinfo{year}{1997}\natexlab{}.
\newblock \showarticletitle{Pragmatism, economics, and the droit moral}.
\newblock \bibinfo{journal}{\emph{NCL Rev.}}  \bibinfo{volume}{76} (\bibinfo{year}{1997}), \bibinfo{pages}{1}.
\newblock


\bibitem[Coyle and Manley(2024)]%
        {coyle2024value}
\bibfield{author}{\bibinfo{person}{Diane Coyle} {and} \bibinfo{person}{Annabel Manley}.} \bibinfo{year}{2024}\natexlab{}.
\newblock \showarticletitle{What is the value of data? A review of empirical methods}.
\newblock \bibinfo{journal}{\emph{Journal of Economic Surveys}} \bibinfo{volume}{38}, \bibinfo{number}{4} (\bibinfo{year}{2024}), \bibinfo{pages}{1317--1337}.
\newblock


\bibitem[DeepMind(2017)]%
        {deepmind}
\bibfield{author}{\bibinfo{person}{DeepMind}.} \bibinfo{year}{2017}\natexlab{}.
\newblock \bibinfo{title}{{Why we launched DeepMind Ethics and Society}}.
\newblock \bibinfo{howpublished}{\url{https://deepmind.google/discover/blog/why-we-launched-deepmind-ethics-society/}}.
\newblock


\bibitem[Demirci et~al\mbox{.}(2023)]%
        {demirci2023ai}
\bibfield{author}{\bibinfo{person}{Ozge Demirci}, \bibinfo{person}{Jonas Hannane}, {and} \bibinfo{person}{Xinrong Zhu}.} \bibinfo{year}{2023}\natexlab{}.
\newblock \showarticletitle{Who Is AI Replacing? The Impact of Genera-tive AI on Online Freelancing Platforms}.
\newblock  (\bibinfo{year}{2023}).
\newblock


\bibitem[Doctorow(2023)]%
        {doctorow2023internet}
\bibfield{author}{\bibinfo{person}{Cory Doctorow}.} \bibinfo{year}{2023}\natexlab{}.
\newblock \bibinfo{booktitle}{\emph{The Internet Con: How to Seize the Means of Computation}}.
\newblock \bibinfo{publisher}{Verso Books}.
\newblock


\bibitem[Economics(2024)]%
        {future_unscripted}
\bibfield{author}{\bibinfo{person}{CVL Economics}.} \bibinfo{year}{2024}\natexlab{}.
\newblock \bibinfo{title}{{Future Unscripted: The Impact of Generative Artificial Intelligence on Entertainment Industry Jobs}}.
\newblock \bibinfo{howpublished}{\url{https://static1.squarespace.com/static/5ce331b47a39b9000198fffa/t/65bacda1c217fc5971cec735/1706741922128/Future+Unscripted+-+The+Impact+of+Generative+Artificial+Intelligence+on+Entertainment+Industry+Jobs+-+pages_compressed}}.
\newblock


\bibitem[Elderkin(2024)]%
        {game_conf}
\bibfield{author}{\bibinfo{person}{Beth Elderkin}.} \bibinfo{year}{2024}\natexlab{}.
\newblock \bibinfo{title}{{GDC 2024 State Of The Game Industry: Devs Discuss Layoffs, Generative AI, And More}}.
\newblock \bibinfo{howpublished}{\url{https://gdconf.com/news/gdc-2024-state-game-industry-devs-discuss-layoffs-generative-ai-and-more}}.
\newblock


\bibitem[Epstein et~al\mbox{.}(2020)]%
        {epstein2020gets}
\bibfield{author}{\bibinfo{person}{Ziv Epstein}, \bibinfo{person}{Sydney Levine}, \bibinfo{person}{David~G Rand}, {and} \bibinfo{person}{Iyad Rahwan}.} \bibinfo{year}{2020}\natexlab{}.
\newblock \showarticletitle{Who gets credit for AI-generated art?}
\newblock \bibinfo{journal}{\emph{Iscience}} \bibinfo{volume}{23}, \bibinfo{number}{9} (\bibinfo{year}{2020}).
\newblock


\bibitem[{European Commission}(2018)]%
        {EUdataregulations2018}
\bibfield{author}{\bibinfo{person}{{European Commission}}.} \bibinfo{year}{2018}\natexlab{}.
\newblock \bibinfo{title}{{2018 {R}eform of {EU} data protection rules}}.
\newblock \bibinfo{howpublished}{Available at \url{https://ec.europa.eu/commission/sites/beta-political/files/data-protection-factsheet-changes_en.pdf}}.
\newblock


\bibitem[Felten et~al\mbox{.}(2021)]%
        {felten2021occupational}
\bibfield{author}{\bibinfo{person}{Edward Felten}, \bibinfo{person}{Manav Raj}, {and} \bibinfo{person}{Robert Seamans}.} \bibinfo{year}{2021}\natexlab{}.
\newblock \showarticletitle{Occupational, industry, and geographic exposure to artificial intelligence: A novel dataset and its potential uses}.
\newblock \bibinfo{journal}{\emph{Strategic Management Journal}} \bibinfo{volume}{42}, \bibinfo{number}{12} (\bibinfo{year}{2021}), \bibinfo{pages}{2195--2217}.
\newblock


\bibitem[Fisk(2006)]%
        {Fisk}
\bibfield{author}{\bibinfo{person}{Catherine~L. Fisk}.} \bibinfo{year}{2006}\natexlab{}.
\newblock \showarticletitle{Credit Where It's Due: The Law and Norms of Attribution}.
\newblock \bibinfo{journal}{\emph{Georgetown Law Journal}} (\bibinfo{year}{2006}).
\newblock


\bibitem[Fleckenstein et~al\mbox{.}(2023)]%
        {fleckenstein2023data}
\bibfield{author}{\bibinfo{person}{Mike Fleckenstein}, \bibinfo{person}{Ali Obaidi}, {and} \bibinfo{person}{Nektaria Tryfona}.} \bibinfo{year}{2023}\natexlab{}.
\newblock \showarticletitle{Data Valuation: Use Cases, Desiderata, and Approaches}. In \bibinfo{booktitle}{\emph{Proceedings of the Second ACM Data Economy Workshop}}. \bibinfo{pages}{48--52}.
\newblock


\bibitem[for Strategic and Studies(2023)]%
        {csis}
\bibfield{author}{\bibinfo{person}{Center for Strategic} {and} \bibinfo{person}{International Studies}.} \bibinfo{year}{2023}\natexlab{}.
\newblock \bibinfo{title}{{AI Regulation is Coming- What is the Likely Outcome?}}
\newblock \bibinfo{howpublished}{\url{https://www.csis.org/blogs/strategic-technologies-blog/ai-regulation-coming-what-likely-outcome}}.
\newblock


\bibitem[for the Protection~of Literary and Works(1886)]%
        {6bisBerne}
\bibfield{author}{\bibinfo{person}{Berne~Convention for the Protection~of Literary} {and} \bibinfo{person}{Artistic Works}.} \bibinfo{year}{1886}\natexlab{}.
\newblock \bibinfo{howpublished}{Art. 6bis}.
\newblock


\bibitem[Getty Images~(US)(2023)]%
        {Getty}
\bibfield{author}{\bibinfo{person}{INC Getty Images~(US), INC v Stability~AI}.} \bibinfo{year}{2023}\natexlab{}.
\newblock \bibinfo{howpublished}{\url{https://cases.justia.com/federal/district-courts/delaware/dedce/1:2023cv00135/81407/1/0.pdf?ts=1682222407}}.
\newblock


\bibitem[Han et~al\mbox{.}(2024b)]%
        {han2024understanding}
\bibfield{author}{\bibinfo{person}{Jiyeon Han}, \bibinfo{person}{Eunseo Yang}, {and} \bibinfo{person}{Uran Oh}.} \bibinfo{year}{2024}\natexlab{b}.
\newblock \showarticletitle{Understanding the Use of AI-Based Audio Generation Models by End-Users}. In \bibinfo{booktitle}{\emph{Extended Abstracts of the CHI Conference on Human Factors in Computing Systems}}. \bibinfo{pages}{1--7}.
\newblock


\bibitem[Han et~al\mbox{.}(2024a)]%
        {10.1145/3613904.3642133}
\bibfield{author}{\bibinfo{person}{Yuanning Han}, \bibinfo{person}{Ziyi Qiu}, \bibinfo{person}{Jiale Cheng}, {and} \bibinfo{person}{RAY LC}.} \bibinfo{year}{2024}\natexlab{a}.
\newblock \showarticletitle{When Teams Embrace AI: Human Collaboration Strategies in Generative Prompting in a Creative Design Task}. In \bibinfo{booktitle}{\emph{Proceedings of the CHI Conference on Human Factors in Computing Systems}} (<conf-loc>, <city>Honolulu</city>, <state>HI</state>, <country>USA</country>, </conf-loc>) \emph{(\bibinfo{series}{CHI '24})}. \bibinfo{publisher}{Association for Computing Machinery}, \bibinfo{address}{New York, NY, USA}, Article \bibinfo{articleno}{176}, \bibinfo{numpages}{14}~pages.
\newblock
\showISBNx{9798400703300}
\urldef\tempurl%
\url{https://doi.org/10.1145/3613904.3642133}
\showDOI{\tempurl}


\bibitem[Heikkil{\"a}(2022)]%
        {heikkila2022artist}
\bibfield{author}{\bibinfo{person}{Melissa Heikkil{\"a}}.} \bibinfo{year}{2022}\natexlab{}.
\newblock \showarticletitle{This artist is dominating AI-generated art. And he’s not happy about it}.
\newblock \bibinfo{journal}{\emph{MIT Technology Review}} \bibinfo{volume}{125}, \bibinfo{number}{6} (\bibinfo{year}{2022}), \bibinfo{pages}{9--10}.
\newblock


\bibitem[Hennink and Kaiser(2022)]%
        {hennink2022sample}
\bibfield{author}{\bibinfo{person}{Monique Hennink} {and} \bibinfo{person}{Bonnie~N Kaiser}.} \bibinfo{year}{2022}\natexlab{}.
\newblock \showarticletitle{Sample sizes for saturation in qualitative research: A systematic review of empirical tests}.
\newblock \bibinfo{journal}{\emph{Social science \& medicine}}  \bibinfo{volume}{292} (\bibinfo{year}{2022}), \bibinfo{pages}{114523}.
\newblock


\bibitem[Hocevar(1979)]%
        {hocevar1979development}
\bibfield{author}{\bibinfo{person}{Dennis Hocevar}.} \bibinfo{year}{1979}\natexlab{}.
\newblock \showarticletitle{The Development of the Creative Behavior Inventory (CBI).}
\newblock  (\bibinfo{year}{1979}).
\newblock


\bibitem[IBM(2018)]%
        {ibm}
\bibfield{author}{\bibinfo{person}{IBM}.} \bibinfo{year}{2018}\natexlab{}.
\newblock \bibinfo{title}{{AI Fairness 360}}.
\newblock \bibinfo{howpublished}{\url{https://aif360.res.ibm.com/}}.
\newblock


\bibitem[IEEE(2017)]%
        {ieee_ais}
\bibfield{author}{\bibinfo{person}{IEEE}.} \bibinfo{year}{2017}\natexlab{}.
\newblock \bibinfo{title}{{Autonomous and Intelligent Syste,s (AIS)}}.
\newblock \bibinfo{howpublished}{\url{https://deepmind.google/discover/blog/why-we-launched-deepmind-ethics-society/}}.
\newblock


\bibitem[Im et~al\mbox{.}(2021)]%
        {im2021yes}
\bibfield{author}{\bibinfo{person}{Jane Im}, \bibinfo{person}{Jill Dimond}, \bibinfo{person}{Melody Berton}, \bibinfo{person}{Una Lee}, \bibinfo{person}{Katherine Mustelier}, \bibinfo{person}{Mark~S Ackerman}, {and} \bibinfo{person}{Eric Gilbert}.} \bibinfo{year}{2021}\natexlab{}.
\newblock \showarticletitle{Yes: Affirmative consent as a theoretical framework for understanding and imagining social platforms}. In \bibinfo{booktitle}{\emph{Proceedings of the 2021 CHI conference on human factors in computing systems}}. \bibinfo{pages}{1--18}.
\newblock


\bibitem[Inie et~al\mbox{.}(2023)]%
        {inie2023designing}
\bibfield{author}{\bibinfo{person}{Nanna Inie}, \bibinfo{person}{Jeanette Falk}, {and} \bibinfo{person}{Steve Tanimoto}.} \bibinfo{year}{2023}\natexlab{}.
\newblock \showarticletitle{Designing participatory ai: Creative professionals’ worries and expectations about generative ai}. In \bibinfo{booktitle}{\emph{Extended Abstracts of the 2023 CHI Conference on Human Factors in Computing Systems}}. \bibinfo{pages}{1--8}.
\newblock


\bibitem[Initiative(2017)]%
        {ccc}
\bibfield{author}{\bibinfo{person}{Cultural Intellectual Property~Rights Initiative}.} \bibinfo{year}{2017}\natexlab{}.
\newblock \bibinfo{title}{{Consent Credit Compensation: The Legal Literacy Campaign}}.
\newblock \bibinfo{howpublished}{\url{https://www.culturalintellectualproperty.com/the-3cs}}.
\newblock


\bibitem[Jiang et~al\mbox{.}(2023a)]%
        {jiang2023ai}
\bibfield{author}{\bibinfo{person}{Harry~H Jiang}, \bibinfo{person}{Lauren Brown}, \bibinfo{person}{Jessica Cheng}, \bibinfo{person}{Mehtab Khan}, \bibinfo{person}{Abhishek Gupta}, \bibinfo{person}{Deja Workman}, \bibinfo{person}{Alex Hanna}, \bibinfo{person}{Johnathan Flowers}, {and} \bibinfo{person}{Timnit Gebru}.} \bibinfo{year}{2023}\natexlab{a}.
\newblock \showarticletitle{AI Art and its Impact on Artists}. In \bibinfo{booktitle}{\emph{Proceedings of the 2023 AAAI/ACM Conference on AI, Ethics, and Society}}. \bibinfo{pages}{363--374}.
\newblock


\bibitem[Jiang et~al\mbox{.}(2023b)]%
        {jiang2023evading}
\bibfield{author}{\bibinfo{person}{Zhengyuan Jiang}, \bibinfo{person}{Jinghuai Zhang}, {and} \bibinfo{person}{Neil~Zhenqiang Gong}.} \bibinfo{year}{2023}\natexlab{b}.
\newblock \showarticletitle{Evading watermark based detection of AI-generated content}. In \bibinfo{booktitle}{\emph{Proceedings of the 2023 ACM SIGSAC Conference on Computer and Communications Security}}. \bibinfo{pages}{1168--1181}.
\newblock


\bibitem[Knight(2022)]%
        {x_AI_ethics}
\bibfield{author}{\bibinfo{person}{Will Knight}.} \bibinfo{year}{2022}\natexlab{}.
\newblock \bibinfo{title}{Elon Musk Has Fired Twitter’s ‘Ethical AI’ Team}.
\newblock
\newblock
\urldef\tempurl%
\url{https://www.wired.com/story/twitter-ethical-ai-team/}
\showURL{%
\tempurl}


\bibitem[Kyi et~al\mbox{.}(2024)]%
        {kyi2024doesn}
\bibfield{author}{\bibinfo{person}{Lin Kyi}, \bibinfo{person}{Abraham Mhaidli}, \bibinfo{person}{Cristiana~Teixeira Santos}, \bibinfo{person}{Franziska Roesner}, {and} \bibinfo{person}{Asia~J Biega}.} \bibinfo{year}{2024}\natexlab{}.
\newblock \showarticletitle{“It doesn’t tell me anything about how my data is used”: User Perceptions of Data Collection Purposes}. In \bibinfo{booktitle}{\emph{Proceedings of the CHI Conference on Human Factors in Computing Systems}}. \bibinfo{pages}{1--12}.
\newblock


\bibitem[Leaffer(2019)]%
        {Leaffer}
\bibfield{author}{\bibinfo{person}{Marshall~A. Leaffer}.} \bibinfo{year}{2019}\natexlab{}.
\newblock \bibinfo{booktitle}{\emph{Understanding Copyright Law (7th edition)}}.
\newblock \bibinfo{publisher}{Carolina Academic Press}.
\newblock


\bibitem[Lee et~al\mbox{.}(2023)]%
        {lee2023talkin}
\bibfield{author}{\bibinfo{person}{Katherine Lee}, \bibinfo{person}{A~Feder Cooper}, {and} \bibinfo{person}{James Grimmelmann}.} \bibinfo{year}{2023}\natexlab{}.
\newblock \showarticletitle{Talkin''Bout AI Generation: Copyright and the Generative-AI Supply Chain}.
\newblock \bibinfo{journal}{\emph{arXiv preprint arXiv:2309.08133}} (\bibinfo{year}{2023}).
\newblock


\bibitem[Lehrman and Linnea Sage~v LOVO(2024)]%
        {Lehrman}
\bibfield{author}{\bibinfo{person}{Paul Lehrman} {and} \bibinfo{person}{INC Linnea Sage~v LOVO}.} \bibinfo{year}{2024}\natexlab{}.
\newblock \bibinfo{howpublished}{\url{https://fingfx.thomsonreuters.com/gfx/legaldocs/dwvkzgknepm/AI\%20VOICEOVER\%20LAWSUIT\%20complaint.pdf}}.
\newblock


\bibitem[Li et~al\mbox{.}(2024a)]%
        {li2024user}
\bibfield{author}{\bibinfo{person}{Jie Li}, \bibinfo{person}{Hancheng Cao}, \bibinfo{person}{Laura Lin}, \bibinfo{person}{Youyang Hou}, \bibinfo{person}{Ruihao Zhu}, {and} \bibinfo{person}{Abdallah El~Ali}.} \bibinfo{year}{2024}\natexlab{a}.
\newblock \showarticletitle{User experience design professionals’ perceptions of generative artificial intelligence}. In \bibinfo{booktitle}{\emph{Proceedings of the CHI Conference on Human Factors in Computing Systems}}. \bibinfo{pages}{1--18}.
\newblock


\bibitem[Li et~al\mbox{.}(2024b)]%
        {li2024value}
\bibfield{author}{\bibinfo{person}{Zhuoyan Li}, \bibinfo{person}{Chen Liang}, \bibinfo{person}{Jing Peng}, {and} \bibinfo{person}{Ming Yin}.} \bibinfo{year}{2024}\natexlab{b}.
\newblock \showarticletitle{The Value, Benefits, and Concerns of Generative AI-Powered Assistance in Writing}. In \bibinfo{booktitle}{\emph{Proceedings of the CHI Conference on Human Factors in Computing Systems}}. \bibinfo{pages}{1--25}.
\newblock


\bibitem[Locke(1690)]%
        {Locke}
\bibfield{author}{\bibinfo{person}{John Locke}.} \bibinfo{year}{1690}\natexlab{}.
\newblock \bibinfo{booktitle}{\emph{In Second Treatise of government}}.
\newblock \bibinfo{publisher}{Princeton University Press}, Chapter Property.
\newblock


\bibitem[Love(2023)]%
        {3Cs}
\bibfield{author}{\bibinfo{person}{James Love}.} \bibinfo{year}{2023}\natexlab{}.
\newblock \bibinfo{title}{{We Need Smart Intellectual Property Laws for Artificial Intelligence}}.
\newblock \bibinfo{howpublished}{\url{https://www.scientificamerican.com/article/we-need-smart-intellectual-property-laws-for-artificial-intelligence/}}.
\newblock


\bibitem[Luccioni et~al\mbox{.}(2024)]%
        {watermarking}
\bibfield{author}{\bibinfo{person}{Sasha Luccioni}, \bibinfo{person}{Yacine Jernite}, \bibinfo{person}{Derek Thomas}, \bibinfo{person}{Emily Witko}, \bibinfo{person}{Ezi Ozoani}, \bibinfo{person}{Josef Fukano}, \bibinfo{person}{Vaibhav Srivastav}, \bibinfo{person}{Brigitte Tousignant}, {and} \bibinfo{person}{Margaret Mitchell}.} \bibinfo{year}{2024}\natexlab{}.
\newblock \bibinfo{title}{{AI Watermaking 101: Tools and Techniques}}.
\newblock \bibinfo{howpublished}{Available at \url{https://huggingface.co/blog/watermarking}}.
\newblock


\bibitem[Macron et~al\mbox{.}(2023)]%
        {macron2023green}
\bibfield{author}{\bibinfo{person}{Emmanuel Macron}, \bibinfo{person}{Mia Mottley}, \bibinfo{person}{Luis In{\'a}cio~Lula da Silva}, \bibinfo{person}{Ursula von~der Leyen}, \bibinfo{person}{Charles Michel}, \bibinfo{person}{Olaf Scholz}, \bibinfo{person}{Fumio Kishida}, \bibinfo{person}{William Ruto}, \bibinfo{person}{Macky Sall}, \bibinfo{person}{Cyril Ramaphosa}, {et~al\mbox{.}}} \bibinfo{year}{2023}\natexlab{}.
\newblock \showarticletitle{A green transition that leaves no one behind’: world leaders release open letter}.
\newblock \bibinfo{journal}{\emph{The Guardian}}  \bibinfo{volume}{20} (\bibinfo{year}{2023}).
\newblock


\bibitem[Main~Sequence et~al\mbox{.}(2024)]%
        {Dudesy}
\bibfield{author}{\bibinfo{person}{Ltd Main~Sequence}, \bibinfo{person}{LLC ORS~v Dudesy}, {and} \bibinfo{person}{ORS}.} \bibinfo{year}{2024}\natexlab{}.
\newblock \bibinfo{howpublished}{\url{https://s3.documentcloud.org/documents/24377081/carlin-lawsuit.pdf}}.
\newblock


\bibitem[Mazzucato(2021)]%
        {mazzucato2021mission}
\bibfield{author}{\bibinfo{person}{Mariana Mazzucato}.} \bibinfo{year}{2021}\natexlab{}.
\newblock \bibinfo{booktitle}{\emph{Mission economy: A moonshot guide to changing capitalism}}.
\newblock \bibinfo{publisher}{Penguin UK}.
\newblock


\bibitem[McDonald et~al\mbox{.}(2019)]%
        {mcdonald2019reliability}
\bibfield{author}{\bibinfo{person}{Nora McDonald}, \bibinfo{person}{Sarita Schoenebeck}, {and} \bibinfo{person}{Andrea Forte}.} \bibinfo{year}{2019}\natexlab{}.
\newblock \showarticletitle{Reliability and inter-rater reliability in qualitative research: Norms and guidelines for CSCW and HCI practice}.
\newblock \bibinfo{journal}{\emph{Proceedings of the ACM on human-computer interaction}} \bibinfo{volume}{3}, \bibinfo{number}{CSCW} (\bibinfo{year}{2019}), \bibinfo{pages}{1--23}.
\newblock


\bibitem[Messeri and Crockett(2024)]%
        {messeri2024artificial}
\bibfield{author}{\bibinfo{person}{Lisa Messeri} {and} \bibinfo{person}{MJ Crockett}.} \bibinfo{year}{2024}\natexlab{}.
\newblock \showarticletitle{Artificial intelligence and illusions of understanding in scientific research}.
\newblock \bibinfo{journal}{\emph{Nature}} \bibinfo{volume}{627}, \bibinfo{number}{8002} (\bibinfo{year}{2024}), \bibinfo{pages}{49--58}.
\newblock


\bibitem[Miernicki and Ng(2021)]%
        {miernicki}
\bibfield{author}{\bibinfo{person}{Martin Miernicki} {and} \bibinfo{person}{Irene Ng}.} \bibinfo{year}{2021}\natexlab{}.
\newblock \showarticletitle{Artificial intelligence and moral rights}.
\newblock \bibinfo{journal}{\emph{Ai \& Society}} \bibinfo{volume}{36}, \bibinfo{number}{1} (\bibinfo{year}{2021}), \bibinfo{pages}{319--329}.
\newblock


\bibitem[Monta(1958)]%
        {monta1958concept}
\bibfield{author}{\bibinfo{person}{Rudolf Monta}.} \bibinfo{year}{1958}\natexlab{}.
\newblock \showarticletitle{The Concept of Copyright Versus the Droit d'Auteur}.
\newblock \bibinfo{journal}{\emph{S. Cal. L. Rev.}}  \bibinfo{volume}{32} (\bibinfo{year}{1958}), \bibinfo{pages}{177}.
\newblock


\bibitem[Mossoff(2002)]%
        {Locke_1}
\bibfield{author}{\bibinfo{person}{Adam Mossoff}.} \bibinfo{year}{2002}\natexlab{}.
\newblock \showarticletitle{Locke’s Labor Lost}.
\newblock \bibinfo{journal}{\emph{University of Chicago Law School Roundtable}} (\bibinfo{year}{2002}).
\newblock


\bibitem[OECD(2021)]%
        {oecd_business}
\bibfield{author}{\bibinfo{person}{OECD}.} \bibinfo{year}{2021}\natexlab{}.
\newblock \bibinfo{title}{{Enterprises by business size}}.
\newblock \bibinfo{howpublished}{\url{https://data.oecd.org/entrepreneur/enterprises-by-business-size.htm}}.
\newblock


\bibitem[of~Privacy~Professionals(2024)]%
        {iapp}
\bibfield{author}{\bibinfo{person}{The International~Association of Privacy~Professionals}.} \bibinfo{year}{2024}\natexlab{}.
\newblock \bibinfo{title}{{Global AI Law and Policy Tracker}}.
\newblock \bibinfo{howpublished}{\url{https://iapp.org/resources/article/global-ai-legislation-tracker/}}.
\newblock


\bibitem[of~the European~Parliament and of~the Council on the legal protection~of databases(1996)]%
        {databasedir}
\bibfield{author}{\bibinfo{person}{Directive~96/9/EC of~the European~Parliament} {and} \bibinfo{person}{of~the Council on the legal protection~of databases}.} \bibinfo{year}{1996}\natexlab{}.
\newblock \bibinfo{howpublished}{Recital 28}.
\newblock


\bibitem[O'Neill et~al\mbox{.}(2018)]%
        {o2018eu}
\bibfield{author}{\bibinfo{person}{Dan O'Neill}, \bibinfo{person}{Herv{\'e} Corvellec}, {et~al\mbox{.}}} \bibinfo{year}{2018}\natexlab{}.
\newblock \showarticletitle{The EU needs a stability and wellbeing pact, not more growth}.
\newblock \bibinfo{howpublished}{\url{https://www.theguardian.com/environment/2023/jun/21/a-green-transition-that-leaves-no-one-behind-world-leaders-release-open-letter}}.
\newblock  (\bibinfo{year}{2018}).
\newblock


\bibitem[Orlitz(2023)]%
        {oritz_art}
\bibfield{author}{\bibinfo{person}{Karla Orlitz}.} \bibinfo{year}{2023}\natexlab{}.
\newblock \bibinfo{title}{{Why AI Models are not inspired like humans.}}
\newblock \bibinfo{howpublished}{\url{https://www.kortizblog.com/blog/why-ai-models-are-not-inspired-like-humans}}.
\newblock


\bibitem[Palani and Ramos(2024)]%
        {palani2024evolving}
\bibfield{author}{\bibinfo{person}{Srishti Palani} {and} \bibinfo{person}{Gonzalo Ramos}.} \bibinfo{year}{2024}\natexlab{}.
\newblock \showarticletitle{Evolving Roles and Workflows of Creative Practitioners in the Age of Generative AI}. In \bibinfo{booktitle}{\emph{Proceedings of the 16th Conference on Creativity \& Cognition}}. \bibinfo{pages}{170--184}.
\newblock


\bibitem[Raworth(2018)]%
        {raworth2018doughnut}
\bibfield{author}{\bibinfo{person}{Kate Raworth}.} \bibinfo{year}{2018}\natexlab{}.
\newblock \bibinfo{booktitle}{\emph{Doughnut economics: Seven ways to think like a 21st century economist}}.
\newblock \bibinfo{publisher}{Chelsea Green Publishing}.
\newblock


\bibitem[Rigamonti(2006)]%
        {rigamonti}
\bibfield{author}{\bibinfo{person}{Cyrill~P Rigamonti}.} \bibinfo{year}{2006}\natexlab{}.
\newblock \showarticletitle{Deconstructing moral rights}.
\newblock \bibinfo{journal}{\emph{Harv. Int'l LJ}}  \bibinfo{volume}{47} (\bibinfo{year}{2006}), \bibinfo{pages}{353}.
\newblock


\bibitem[Robertson(2024)]%
        {grok}
\bibfield{author}{\bibinfo{person}{Adi Robertson}.} \bibinfo{year}{2024}\natexlab{}.
\newblock \bibinfo{title}{X’s new AI image generator will make anything from Taylor Swift in lingerie to Kamala Harris with a gun}.
\newblock
\newblock
\urldef\tempurl%
\url{https://www.theverge.com/2024/8/14/24220173/xai-grok-image-generator-misinformation-offensive-imges}
\showURL{%
\tempurl}


\bibitem[Shan et~al\mbox{.}(2023)]%
        {shan2023glaze}
\bibfield{author}{\bibinfo{person}{Shawn Shan}, \bibinfo{person}{Jenna Cryan}, \bibinfo{person}{Emily Wenger}, \bibinfo{person}{Haitao Zheng}, \bibinfo{person}{Rana Hanocka}, {and} \bibinfo{person}{Ben~Y Zhao}.} \bibinfo{year}{2023}\natexlab{}.
\newblock \showarticletitle{Glaze: Protecting artists from style mimicry by $\{$Text-to-Image$\}$ models}. In \bibinfo{booktitle}{\emph{32nd USENIX Security Symposium (USENIX Security 23)}}. \bibinfo{pages}{2187--2204}.
\newblock


\bibitem[Shan et~al\mbox{.}(2024)]%
        {shan2024nightshade}
\bibfield{author}{\bibinfo{person}{Shawn Shan}, \bibinfo{person}{Wenxin Ding}, \bibinfo{person}{Josephine Passananti}, \bibinfo{person}{Stanley Wu}, \bibinfo{person}{Haitao Zheng}, {and} \bibinfo{person}{Ben~Y Zhao}.} \bibinfo{year}{2024}\natexlab{}.
\newblock \showarticletitle{Nightshade: Prompt-Specific Poisoning Attacks on Text-to-Image Generative Models}. In \bibinfo{booktitle}{\emph{2024 IEEE Symposium on Security and Privacy (SP)}}. IEEE Computer Society, \bibinfo{pages}{212--212}.
\newblock


\bibitem[Silverman et~al\mbox{.}(2023)]%
        {Silverman}
\bibfield{author}{\bibinfo{person}{Sarah Silverman}, \bibinfo{person}{ORS v OpenAI}, {and} \bibinfo{person}{ORS}.} \bibinfo{year}{2023}\natexlab{}.
\newblock \bibinfo{howpublished}{\url{https://news.justia.com/wp-content/uploads/2023/07/Silverman-et-al.-v.-OpenAI-Complaint.pdf}}.
\newblock


\bibitem[Singh(2022)]%
        {singh2022consent}
\bibfield{author}{\bibinfo{person}{Munindar~P Singh}.} \bibinfo{year}{2022}\natexlab{}.
\newblock \showarticletitle{Consent as a foundation for responsible autonomy}. In \bibinfo{booktitle}{\emph{Proceedings of the AAAI Conference on Artificial Intelligence}}, Vol.~\bibinfo{volume}{36}. \bibinfo{pages}{12301--12306}.
\newblock


\bibitem[Sprigman et~al\mbox{.}(2013)]%
        {Sprigman}
\bibfield{author}{\bibinfo{person}{Christopher~Jon Sprigman}, \bibinfo{person}{Christopher Buccafusco}, {and} \bibinfo{person}{Zachary Burns}.} \bibinfo{year}{2013}\natexlab{}.
\newblock \showarticletitle{What's a Name Worth: Experimental Tests of the Value of Attribution in Intellectual Property}.
\newblock \bibinfo{journal}{\emph{Boston University Law Review}} (\bibinfo{year}{2013}).
\newblock


\bibitem[Stempel(2024)]%
        {nvdia_sued}
\bibfield{author}{\bibinfo{person}{Jonathan Stempel}.} \bibinfo{year}{2024}\natexlab{}.
\newblock \bibinfo{title}{{Nvidia is sued by authors over AI use of copyrighted works}}.
\newblock \bibinfo{howpublished}{\url{https://www.reuters.com/technology/nvidia-is-sued-by-authors-over-ai-use-copyrighted-works-2024-03-10/}}.
\newblock


\bibitem[Stiglitz(2009)]%
        {stiglitz2009report}
\bibfield{author}{\bibinfo{person}{Joseph~E Stiglitz}.} \bibinfo{year}{2009}\natexlab{}.
\newblock \showarticletitle{Report by the Commission on the Measurement of Economic Performance and social Progress}.
\newblock  (\bibinfo{year}{2009}).
\newblock


\bibitem[Stone et~al\mbox{.}(2016)]%
        {stone2022artificial}
\bibfield{author}{\bibinfo{person}{Peter Stone}, \bibinfo{person}{Rodney Brooks}, \bibinfo{person}{Erik Brynjolfsson}, \bibinfo{person}{Ryan Calo}, \bibinfo{person}{Oren Etzioni}, \bibinfo{person}{Greg Hager}, \bibinfo{person}{Julia Hirschberg}, \bibinfo{person}{Shivaram Kalyanakrishnan}, \bibinfo{person}{Ece Kamar}, \bibinfo{person}{Sarit Kraus}, {et~al\mbox{.}}} \bibinfo{year}{2016}\natexlab{}.
\newblock \showarticletitle{Artificial intelligence and life in 2030: the one hundred year study on artificial intelligence}.
\newblock \bibinfo{journal}{\emph{Report, Stanford University}} (\bibinfo{year}{2016}).
\newblock


\bibitem[Times(2023)]%
        {hollywood_writers}
\bibfield{author}{\bibinfo{person}{Los~Angeles Times}.} \bibinfo{year}{2023}\natexlab{}.
\newblock \bibinfo{title}{{Writers’ strike: What happened, how it ended and its impact on Hollywood}}.
\newblock \bibinfo{howpublished}{\url{https://www.latimes.com/entertainment-arts/business/story/2023-05-01/writers-strike-what-to-know-wga-guild-hollywood-productions}}.
\newblock


\bibitem[Tushnet(2007)]%
        {tushnet2007payment}
\bibfield{author}{\bibinfo{person}{Rebecca Tushnet}.} \bibinfo{year}{2007}\natexlab{}.
\newblock \showarticletitle{Payment in credit: Copyright law and subcultural creativity}.
\newblock \bibinfo{journal}{\emph{Law and contemporary problems}} \bibinfo{volume}{70}, \bibinfo{number}{2} (\bibinfo{year}{2007}), \bibinfo{pages}{135--174}.
\newblock


\bibitem[v~Microsoft~Corporation and ORS(2023)]%
        {NYT}
\bibfield{author}{\bibinfo{person}{The New York Times Company~(NYT) v Microsoft~Corporation} {and} \bibinfo{person}{ORS}.} \bibinfo{year}{2023}\natexlab{}.
\newblock \bibinfo{howpublished}{\url{https://nytco-assets.nytimes.com/2023/12/NYT_Complaint_Dec2023.pdf}}.
\newblock


\bibitem[Vincent(2022)]%
        {githubcopilot}
\bibfield{author}{\bibinfo{person}{James Vincent}.} \bibinfo{year}{2022}\natexlab{}.
\newblock \bibinfo{title}{{The lawsuit that could rewrite the rules of AI copyright}}.
\newblock \bibinfo{howpublished}{\url{https://www.theverge.com/2022/11/8/23446821/microsoft-openai-github-copilot-class-action-lawsuit-ai-copyright-violation-training-data}}.
\newblock
\newblock
\shownote{[Online; accessed 07-March-2024]}.


\bibitem[Vincent and Hecht(2024)]%
        {vincent2024data}
\bibfield{author}{\bibinfo{person}{Nicholas Vincent} {and} \bibinfo{person}{Brent Hecht}.} \bibinfo{year}{2024}\natexlab{}.
\newblock \showarticletitle{Sharing the Winnings of AI with Data Dividends: Challenges with “Meritocratic” Data Valuation}. In \bibinfo{booktitle}{\emph{EEAMO Workshop}}.
\newblock


\bibitem[Walsh(2023)]%
        {legal_genAI}
\bibfield{author}{\bibinfo{person}{Dylan Walsh}.} \bibinfo{year}{2023}\natexlab{}.
\newblock \bibinfo{title}{{The legal issues presented by generative AI}}.
\newblock \bibinfo{howpublished}{\url{https://mitsloan.mit.edu/ideas-made-to-matter/legal-issues-presented-generative-ai}}.
\newblock


\bibitem[Watercutter and Bedingfield(2023)]%
        {hollywood_actors}
\bibfield{author}{\bibinfo{person}{Angela Watercutter} {and} \bibinfo{person}{Will Bedingfield}.} \bibinfo{year}{2023}\natexlab{}.
\newblock \bibinfo{title}{{Hollywood Actors Strike Ends With a Deal That Will Impact AI and Streaming for Decades}}.
\newblock \bibinfo{howpublished}{\url{https://www.wired.com/story/hollywood-actors-strike-ends-ai-streaming/}}.
\newblock


\bibitem[Widder et~al\mbox{.}(2023)]%
        {widder2023s}
\bibfield{author}{\bibinfo{person}{David~Gray Widder}, \bibinfo{person}{Derrick Zhen}, \bibinfo{person}{Laura Dabbish}, {and} \bibinfo{person}{James Herbsleb}.} \bibinfo{year}{2023}\natexlab{}.
\newblock \showarticletitle{It’s about power: What ethical concerns do software engineers have, and what do they (feel they can) do about them?}. In \bibinfo{booktitle}{\emph{Proceedings of the 2023 ACM Conference on Fairness, Accountability, and Transparency}}. \bibinfo{pages}{467--479}.
\newblock


\bibitem[Woodruff et~al\mbox{.}(2024)]%
        {woodruff2024knowledge}
\bibfield{author}{\bibinfo{person}{Allison Woodruff}, \bibinfo{person}{Renee Shelby}, \bibinfo{person}{Patrick~Gage Kelley}, \bibinfo{person}{Steven Rousso-Schindler}, \bibinfo{person}{Jamila Smith-Loud}, {and} \bibinfo{person}{Lauren Wilcox}.} \bibinfo{year}{2024}\natexlab{}.
\newblock \showarticletitle{How Knowledge Workers Think Generative AI Will (Not) Transform Their Industries}.
\newblock \bibinfo{journal}{\emph{CHI 2024}} (\bibinfo{year}{2024}).
\newblock


\bibitem[Wu et~al\mbox{.}(2023)]%
        {wu2023not}
\bibfield{author}{\bibinfo{person}{Yankun Wu}, \bibinfo{person}{Yuta Nakashima}, {and} \bibinfo{person}{Noa Garcia}.} \bibinfo{year}{2023}\natexlab{}.
\newblock \showarticletitle{Not only generative art: Stable diffusion for content-style disentanglement in art analysis}. In \bibinfo{booktitle}{\emph{Proceedings of the 2023 ACM International conference on multimedia retrieval}}. \bibinfo{pages}{199--208}.
\newblock


\bibitem[Yurrita et~al\mbox{.}(2022)]%
        {yurrita}
\bibfield{author}{\bibinfo{person}{Mireia Yurrita}, \bibinfo{person}{Dave Murray-Rust}, \bibinfo{person}{Agathe Balayn}, {and} \bibinfo{person}{Alessandro Bozzon}.} \bibinfo{year}{2022}\natexlab{}.
\newblock \showarticletitle{Towards a multi-stakeholder value-based assessment framework for algorithmic systems}. In \bibinfo{booktitle}{\emph{Proceedings of the 2022 ACM Conference on Fairness, Accountability, and Transparency}}. \bibinfo{pages}{535--563}.
\newblock


\bibitem[Zou et~al\mbox{.}(2018)]%
        {zou2018ve}
\bibfield{author}{\bibinfo{person}{Yixin Zou}, \bibinfo{person}{Abraham~H Mhaidli}, \bibinfo{person}{Austin McCall}, {and} \bibinfo{person}{Florian Schaub}.} \bibinfo{year}{2018}\natexlab{}.
\newblock \showarticletitle{" I've Got Nothing to Lose": Consumers' Risk Perceptions and Protective Actions after the Equifax Data Breach}. In \bibinfo{booktitle}{\emph{Fourteenth Symposium on Usable Privacy and Security (SOUPS 2018)}}. \bibinfo{pages}{197--216}.
\newblock


\end{thebibliography}

\appendix
\section{\\Interview Script}
\label{sec:appendixa}

\textbf{Preamble:}

``Hello, my name is [researcher’s name]. I am a [researcher’s position] working with [research
team’s affiliation]. Today, I will be conducing an interview which should take about an hour
long. The purpose of this interview is to see what creative professionals such as artists,
designers, writers, and programmers think about generative AI, and how companies can be
more ethical when handling creative data.

This study is for research purposes only, and you only need to share the extent to which you are
comfortable sharing. In the data analysis and reporting of results, your identity and
company, publisher, and/or clients will be anonymized.

Before we begin, do you have any questions?''

\noindent \textbf{Consent and demographics:}

``In the Zoom chat, I will send a link to a consent form for being audio recorded. Your consent is optional. If you consent, the audio recording will only be used for research purposes and for further data analysis; your identity will be anonymized. If you do not consent, which is fine as well, I will be taking notes during our interview instead of audio recording.

When you have completed the consent form, please fill out the demographics questionnaire on
the next page of the link.''

[Researcher sends consent form and demographics questionnaire in Zoom chat]

\noindent \textbf{Demographics form:}
\begin{enumerate}
    \item Age
    \begin{itemize}
        \item Fill in a number, or
        \item Prefer not to disclose
    \end{itemize}
    \item Gender
    \begin{itemize}
        \item Female,
        \item Male,
        \item Non-binary, or
        \item Prefer not to disclose
    \end{itemize}
    \item Where are you located?
    \begin{itemize}
        \item Country drop-down list, or
        \item Prefer not to disclose
    \end{itemize}
    \item What kind of creative domain(s) are you professionally working in?
    \begin{itemize}
        \item Visual art or design,
        \item Writing, and/or
        \item Programming
    \end{itemize}
    \item How long have you been doing this creative domain professionally? [select all that apply]
    \begin{itemize}
        \item Fill in a time period
    \end{itemize}
    \item Who do you typically work with? [select all that apply]
    \begin{itemize}
        \item Employed by a company,
        \item Working with publishers, 
        \item Freelancer and selling my work on gig platforms, and/or
        \item Other: [fill in]
    \end{itemize}
    \item How long have you been working with your selection above?
    \begin{itemize}
        \item Fill in a number
    \end{itemize}
\end{enumerate}

[Researcher begins audio recording if participant consented]

\textbf{Interview begins:}

\textit{First, I want to ask some background questions:}
\begin{enumerate}
    \item Can you tell me a bit about your background and creative career?
    \item What are your thoughts of, and experiences with generative AI relating to your creative work?
\end{enumerate}

\textit{Here, I am interested in hearing about how your company, platform, and/or publisher is dealing with generative AI:}
\begin{enumerate}[resume]
    \item If you don't mind sharing, what has your company, platform, and/or publisher done in relation to generative AI and creative work?
    \item To the extent you are allowed, or are comfortable sharing, how has your company, platform, and/or publisher gone about setting up rules about how to use (or not use) generative AI?
    \item Were you informed by your company, platform, and/or publisher about rules changing due to generative AI?
\end{enumerate}

\textit{Now, I have some questions about ownership and hypothetical questions about if your work were used to train AI models:}
\begin{enumerate}[resume]
    \item How much ownership do you want over your creative work while working for a company, platform, and/or publisher?
    \begin{enumerate}
        \item How do your opinions of ownership change with generative AI in the picture?
    \end{enumerate}
    \item How would you want your creative work to be handled once you leave that company, platform, and/or publisher?
    \item Would you be fine if your work were used to train AI models, hypothetically speaking?
    \begin{enumerate}
        \item (If yes) Is being notified that your work is being used to train an AI model important to you?
        \begin{itemize}
            \item  (If yes) How would you like to be notified that your work is being used to train AI models?
        \end{itemize}
        \item (If yes) Would you like to be attributed or prefer to be anonymous for your work?
        \begin{itemize}
            \item (Follow up, depending on response) Why would you prefer to be attributed/anonymous?
        \end{itemize}
        \item (If yes) How important is compensation if your work were to be used to train AI models?
        \begin{itemize}
            \item Does your employment status with a company, platform, and/or publisher matter for compensation?
        \end{itemize}
    \end{enumerate}
\end{enumerate}

\textit{Now, I have some questions about working conditions at your company/publisher/platform:}
\begin{enumerate}[resume]
    \item When changes are made to your working conditions, are you usually notified?
    \begin{enumerate}
        \item How would you prefer to be notified?
    \end{enumerate}
    \item Are you part of a union or collective organization?
    \begin{enumerate}
        \item (How) do you communicate with others in your field to discuss working conditions?
    \end{enumerate}
\end{enumerate}

\textit{Here, I will ask about future implications of generative AI:}

\begin{enumerate}[resume]
    \item What are your thoughts about the likely or possible implications/consequences of generative AI for your creative industry?
    \begin{enumerate}
        \item How has generative AI changed how you view your industry?
    \end{enumerate}
    \item How do you think the quality of creative work in your discipline will be impacted by AI?
    \begin{enumerate}
        \item What are some differences you think are present between AI-made vs. human-made work in your field?
        \item What elements indicate that it is generated by an AI vs. a human in your domain?
    \end{enumerate}
    \item How might AI disrupt the usual creative process?
    \item Do you have any concerns about AI replacing your creative work?
    \item What can companies do to deal with creative work more ethically in relation to AI?
\end{enumerate}

[Recording ends, researcher thanks and compensates participant]

\end{document}